\documentclass[aps,prfluids,reprint,longbibliography]{revtex4-2}
\usepackage{amsmath}
\usepackage{graphicx}
\usepackage{hyperref}
\usepackage{xcolor}
\usepackage[normalem]{ulem}
\usepackage{float}

\def\Uv{\mathbf{U}}
\def\uv{\mathbf{u}}

\def\C{\bm{\mathsf{C}}}

\def\Fv{\mathbf{F}}
\def\Q{\mathbf{Q}}

\def\nablav{\boldsymbol{\nabla}}

\def\Uv{\boldsymbol{U}}

\def\C{{\bf C}}

\def\Fv{\mathbf{F}}

\begin{document}

\title{Statistical State Dynamics Eigenmodes and Equilibria support Couette Turbulence}
\author{ Brian F. Farrell}
\affiliation{Department of Earth and Planetary Sciences, Harvard University, Cambridge, U.S.A.}
\author{Petros J. Ioannou}
\affiliation{Department of Physics, National and Kapodistrian University of Athens, Athens, Greece}
\affiliation{Department of Earth and Planetary Sciences, Harvard University, Cambridge, U.S.A.}

\date{\today}

\begin{abstract}

Wide-channel Couette (WCC) turbulence consists primarily of quasi-steady roll–streak structures (RSS) maintained by a self-sustaining process (SSP), yet the Navier–Stokes equations in velocity variables admit no linear RSS instability or stable equilibrium, leaving the WCC turbulent state's analytic basis obscure. We show that, in the statistical state dynamics (SSD) of a second-order closure of the Navier-Stokes equations, Couette turbulence arises from a modal RSS instability and equilibrates as a stable fixed point at spanwise wavenumber 3. This fixed point comprises a streamwise-mean flow and a pair of counterpropagating neutral eigenmodes, regularized by roll advection rather than viscosity. This stable equilibrium identifies both a canonical turbulence solution and its self-sustaining process (SSP), and characterizes their analytic structure.  WCC turbulence arises as a spanwise tiling by this stable fixed-point RSS unit cell. Predictions of this SSD theory are corroborated by comparison with direct numerical simulation.

\end{abstract}

\maketitle

In wall-bounded shear flows, turbulence paradoxically exhibits organized, macroscopic order. Both laboratory experiments \cite{Bech-1994, Papavasiliou-1997, Tillmark-1995, Tillmark-1998, Kitoh-2005, Kitoh-2008}, direct numerical simulations (DNS) and large-eddy simulations (LES) of wide-channel Couette (WCC) turbulence across a broad range of Reynolds numbers have identified a robust equilibrium state dominated by large-scale, streamwise-oriented
roll-streak structures (RSS)
\cite{Komminaho-etal-1996, Tsukahara-etal-2006, Pirozzoli-etal-2014, Avsarkisov-etal-2014,
Rawat-etal-2015,Lee-Moser-2018,Cheng-Pullin-2022, Hoyas-Oberlack-2024}. These structures, which underlie the fundamental dynamics of the turbulent state, exhibit extensive spatial coherence with a streamwise wavenumber near zero and a tightly bound spanwise wavenumber.  As the spanwise aspect ratio increases, the statistical variability of these structures diminishes \cite{Pirozzoli-etal-2014}, suggesting that WCC is converging toward a statistical fixed-point. This fixed-point turbulence offers a unique opportunity to study the essential dynamics of wall-bounded turbulence.

\begin{figure}
    \centering
    \includegraphics[width=1\linewidth]{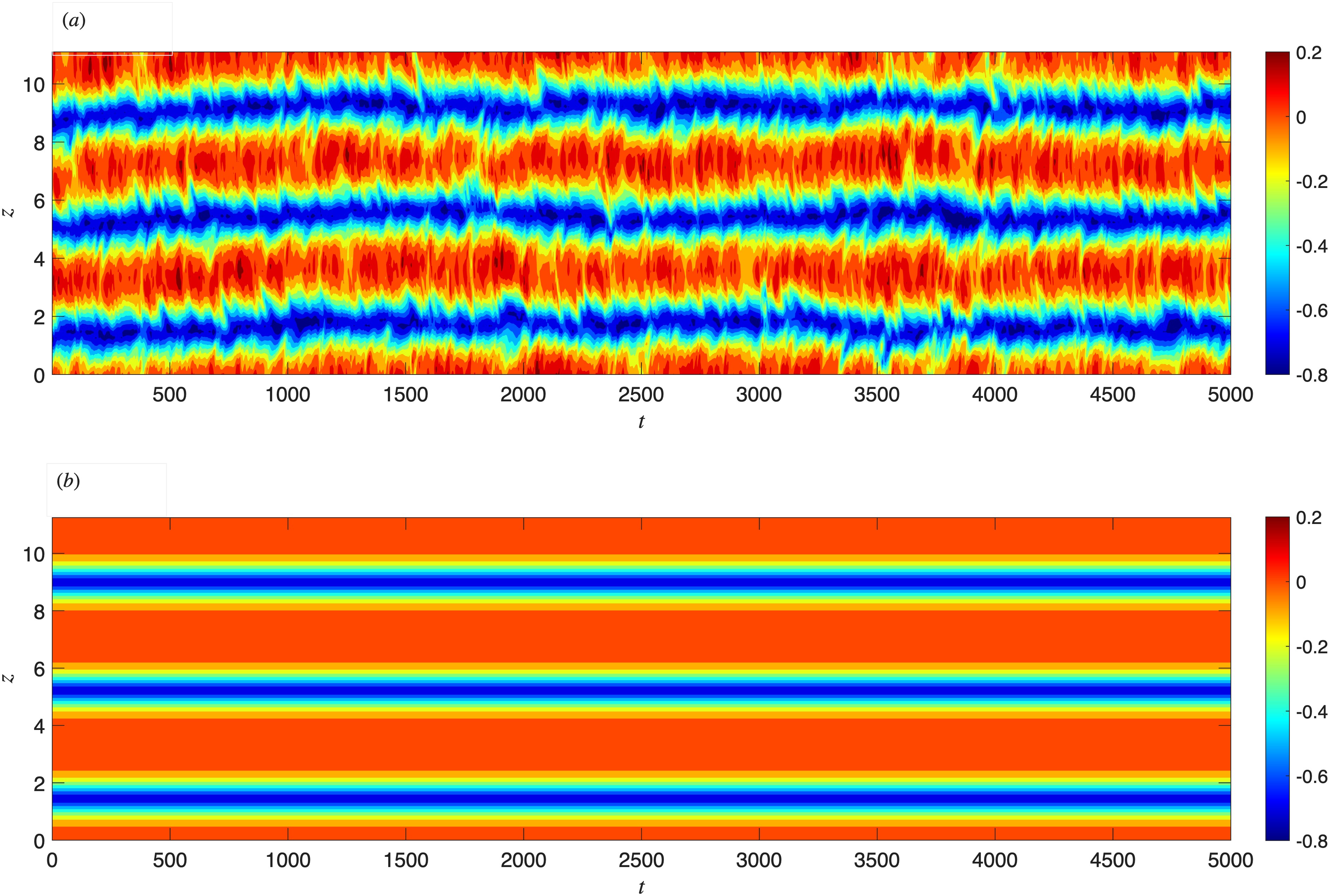}
    \caption{Streamwise velocity as a function of time $t$ and spanwise coordinate $z$ at $y=-0.7$, showing the long-term persistence of the large-scale streaks. \textbf{(a)} DNS of pCf turbulence. \textbf{(b)} The corresponding  S3T with $\Q=0$. Geometry $L_x \times L_z = 1.75\pi \times 3.6\pi$ at $R=600$.}
    \label{fig:Hov}
    \vskip-0.2in
\end{figure}

In this Letter, we demonstrate that the statistical state of WCC turbulence corresponds to a stable fixed-point equilibrium of a closure of the Navier-Stokes (NS) statistical state dynamics (SSD) in a periodic Couette channel with spanwise wavenumber 3. We further show how this stable equilibrium forms the basis for a spanwise tiling of WCC turbulence. Because both this stable equilibrium and the unstable mode from which it arises lack an  analytic characterization in the Navier-Stokes equations (NSE) formulated in traditional velocity variables, they have previously escaped identification despite the striking evidence of fixed-point structure in both observations and simulations of WCC turbulence.

We obtain an analytic characterization of the RSS and the self-sustaining process (SSP) that maintains WCC  turbulence through a four-step procedure:\\
1. Expressing the NSE in statistical state variables.\\
2. Closing the resulting cumulant dynamics at second~ order to derive the SSD.\\
3. Perturbing this SSD linearly to identify the RSS instability.\\
4. Tracking this instability into the nonlinear regime to locate the  finite-amplitude tiling stable equilibrium.

Together, these steps yield an analytic framework that rationalizes the SSP mechanism captured numerically in the DNS.
This second-order SSD closure  is  known as Stochastic Structural Stability Theory (S3T). In this closure, the fluctuation covariance is a primary state variable. Linear perturbation of the covariance matrix allows linear analysis to be brought to bear on the nonlinear instability that underlies the turbulent state. The S3T SSD formulation retains the essential dynamics of wall-bounded turbulence with maximal clarity and simplicity \cite{Farrell-Ioannou-2003-structural,Farrell-Ioannou-2012}.  S3T has succeeded previously in identifying  SSD instabilities  as well  as fixed-point SSD solutions in other turbulent shear flows \cite{Farrell-Ioannou-2007-structure,*Farrell-Ioannou-2009-equatorial,*Farrell-Ioannou-2009-plasmas,*Farrell-Ioannou-2017-Saturn, *Parker-Krommes-2013,*Parker-Krommes-2014-generation,*Bakas-Ioannou-2011,*Bakas-Ioannou-2014-jfm,*Constantinou-etal-2016}.

The motion of an incompressible fluid with constant density is governed by the non-dimensional Navier-Stokes equations,
\begin{equation}
\label{eq:NS}\partial_t \mathbf{u} + (\mathbf{u} \cdot \nabla) \mathbf{u} = -\nabla p + \frac{1}{R} \Delta \mathbf{u}, \quad \nabla \cdot \mathbf{u} = 0,
\end{equation}
where $\mathbf{u}(x, y, z, t)$ and $p$ are the velocity and pressure fields.  The flow is confined between parallel planes at $y = \pm 1$, with periodic boundary conditions in the streamwise ($x$) and spanwise ($z$) directions over lengths $L_x$ and $L_z$.
Variables are non-dimensionalized by the wall speed $U_w$ and the channel half-width $h$, so that the Reynolds number is $R=U_w h/\nu$,
with $\nu$ the kinematic viscosity. As is appropriate for plane Couette flow (pCf), the walls move in opposite directions and we impose no-slip boundary conditions $\mathbf{u}(x,\pm 1,z,t)=\pm\hat{\mathbf{e}}_x$.  The equations are solved numerically; numerical methods and parameters are given in ~\cite{Methods-Supplement}.

In contrast to the spatio-temporal disorder of the RSS in plane Poiseuille flow (pPf) turbulence, the RSSs in WCC turbulence are strongly aligned in the streamwise direction, nearly periodic in the spanwise direction, and exhibit minimal temporal variability. Consequently, they are well represented by the streamwise-averaged velocity field $\langle\mathbf{u}\rangle_x \equiv (U,V,W)$, where $\langle \cdot \rangle_x$ denotes a spatial average over the streamwise coordinate $x$. The non-divergent cross-stream and spanwise velocities, $(V,W)$, constitute primarily the roll component of the RSS and can be expressed using a streamfunction $\Psi(y,z,t)$ such that $(V,W) = (-\partial_z \Psi, \partial_y \Psi)$. The cross-stream velocity $V$ advects the mean streamwise momentum to produce the spanwise streak component of the RSS, defined as $U_s(y, z, t) = U - \langle U \rangle_z$.

In DNS of WCC flow, a turbulent state dominated by RSS structures with characteristic spanwise wavenumber and minimal temporal variability emerges and persists. In contrast, spanwise-confined domains with periodic boundaries exhibit time-dependent dynamics, as seen in DNS of minimal-channel flows \cite{Hamilton-etal-1995}. Yet, interspersed among these time-dependent regimes are discrete windows in channel width for which fixed-point turbulent states exist. A prominent example is the channel with $L_x \times L_z = 1.75\pi \times 3.6\pi$—notably three times the spanwise extent of a canonical minimal channel \cite{Hamilton-etal-1995}. In an interval around this specific channel width the fundamental dynamical equilibrium structure that underlies WCC turbulence is supported.  The spatio-temporal coherence of the RSS in pCf in this domain at $R=600$ is illustrated in Fig. \ref{fig:Hov} (upper panel). Provided the spanwise extent is sufficiently great, this organization remains remarkably persistent over a wide range of channel widths and Reynolds numbers \cite{Pirozzoli-etal-2014, Avsarkisov-etal-2014,Rawat-etal-2015, Lee-Moser-2018,Cheng-Pullin-2022, Hoyas-Oberlack-2024}.

\begin{figure}
    \centering
    \includegraphics[width=1\linewidth]{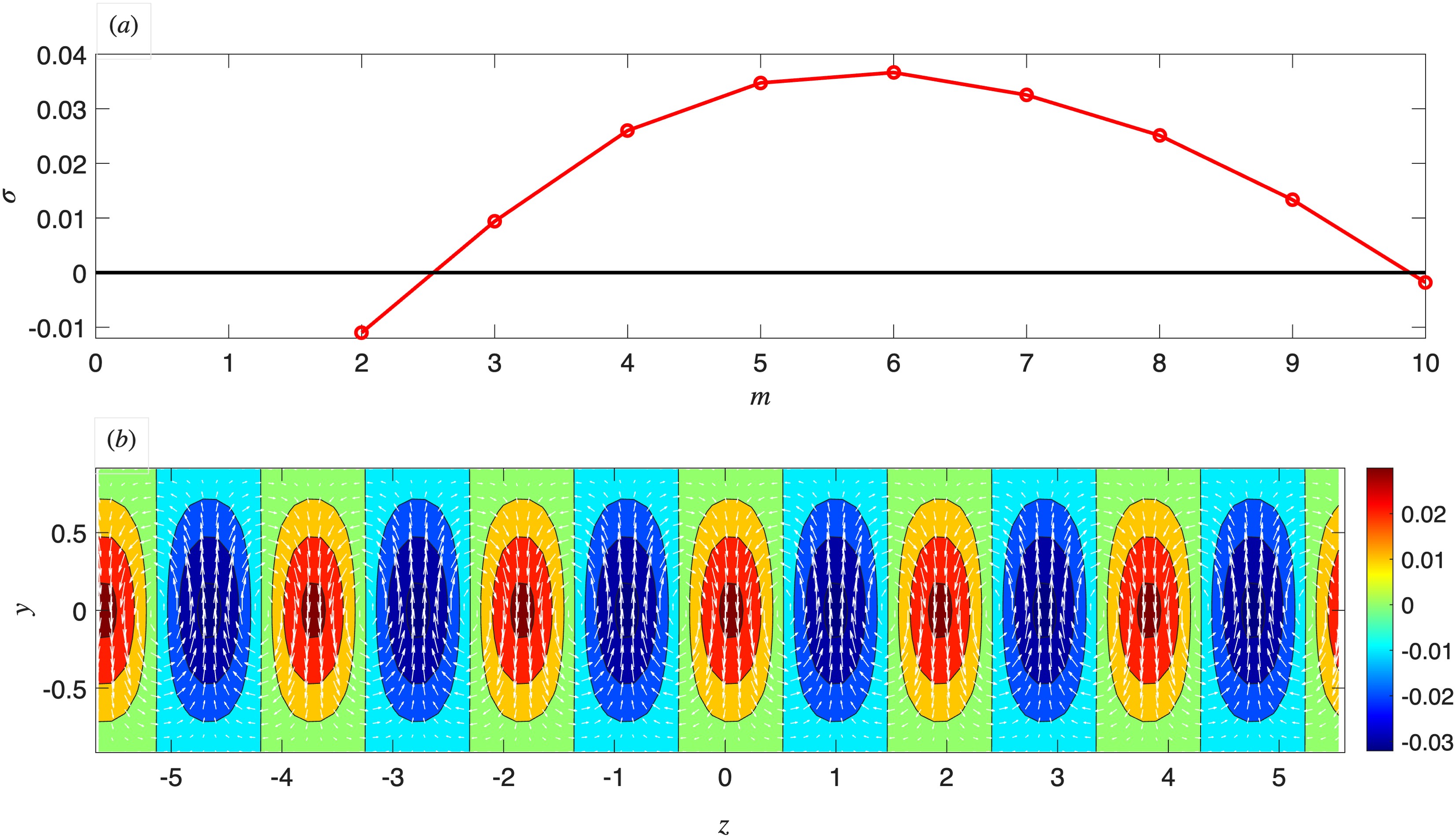}
   \caption{\textbf{(a)} Growth rate of the most unstable S3T eigenmode as a
function of spanwise mode number $m$, shown for the gravest streamwise
wavenumber $k_x = 2\pi/L_x$, at which the instability is maximal; peak
growth occurs at $m=6$. $\Q$
maintains  fluctuation energy density  $0.56\%$ of the
laminar-flow energy density at each $k_x$.
\textbf{(b)} 
$\delta\mathbf{U}$ of the maximally growing eigenmode: contours of
$\delta U$ with cross-flow vectors $(\delta V,\delta W)$; maximum
amplitudes are in the ratio
$[\delta U,\delta V,\delta W]=[1,0.09,0.04]$. Here
$L_x \times L_z = 1.75\pi \times 3.6\pi$ and $R=600$.}
    \label{fig:S3Tstab}
   \vskip-0.2in
\end{figure}

To formulate an analytic theory for the emergence of these roll-streak structures from an unstable eigenmode of the turbulent state, we decompose the velocity and pressure fields into streamwise-mean and fluctuating components: $\mathbf{u} = \mathbf{U} + \mathbf{u}'$ and $p = P + p'$, where $\mathbf{U} = \langle \mathbf{u} \rangle_x$. The system state is then defined by the mean flow $\mathbf{U}$ and the equal-time, two-point fluctuation covariance tensor $\mathbf{C}(\mathbf{x}_1,\mathbf{x}_2,t) = \overline{\mathbf{u}'(\mathbf{x}_1,t) \otimes \mathbf{u}'(\mathbf{x}_2,t)}$, where the overbar denotes an ensemble average. Streamwise averaging the Navier-Stokes equations yields the evolution equation for the mean flow:

\begin{equation}\label{eq:S3Tm}
\partial_t \mathbf{U} + \mathbf{U} \cdot \nabla \mathbf{U} + \nabla P - \frac{1}{R} \Delta \mathbf{U} = \mathcal{L}(\mathbf{C}), \quad \nabla \cdot \mathbf{U} = 0,
\end{equation}
where the Reynolds stress divergence, $\mathcal{L}(\mathbf{C}) = -\langle \mathbf{u}' \cdot \nabla \mathbf{u}' \rangle_x$, is expressed as a linear operator acting on the covariance matrix $\mathbf{C}$. This step invokes the ergodic hypothesis, equating the streamwise spatial average with the statistical ensemble average. To close the system, third-order cumulants are parameterized as a temporally white stochastic excitation with spatial covariance $\mathbf{Q}$. The resulting evolution of the fluctuation covariance in matrix form is governed by the Lyapunov equation:
\begin{equation}\label{eq:C}
\partial_t \mathbf{C} = \mathcal{A}(\mathbf{U}) \mathbf{C} + \mathbf{C} \mathcal{A}^\dagger(\mathbf{U}) + \mathbf{Q},
\end{equation}
where $\mathcal{A}(\mathbf{U})$ is the linear matrix operator governing the evolution of  fluctuations $\mathbf{u}'$. Equations \eqref{eq:S3Tm} and \eqref{eq:C} constitute the S3T statistical state dynamics. By shifting the state variables to the statistical pair $(\mathbf{U}, \mathbf{C})$, S3T transforms the representation of the central nonlinear RSS instability  making it  accessible by linear stability theory. The DNS confirms the existence, growth rate and structure of the analytically characterized S3T instability \cite{Farrell-Ioannou-2017-bifur,Nikolaidis-2024}. In this way, S3T analysis identifies a
pathway to Couette turbulence mediated by an explicit modal instability\cite{Farrell-Ioannou-2012,Farrell-Ioannou-2017-bifur}.

Consider the S3T stability problem for Couette flow in the presence of a low-intensity, spanwise-homogeneous  fluctuation covariance,
$\C_e$,  maintained by $\mathbf{Q}$.  
The associated Reynolds stresses sustain an equilibrium mean profile, $\mathbf{U}_e = (U_e(y), 0, 0)$, that is slightly modified from the laminar Couette solution. Together, $(\mathbf{U}_e, \mathbf{C}_e)$ define an equilibrium  S3T state satisfying the time-independent  Eqs. \eqref{eq:S3Tm} and \eqref{eq:C}.  Perturbations $(\mathbf{\delta U}, \mathbf{\delta C})$ about this equilibrium evolve according to the linearized S3T equations:
\begin{subequations}
\label{eq:S3Tp}
\begin{align}
\partial_t \mathbf{\delta U} & = - \mathbf{U}_e \cdot \nabla \mathbf{\delta U} - \mathbf{\delta U} \cdot \nabla \mathbf{U}_e - \nabla \delta P \nonumber \\ & \quad + R^{-1} \Delta \mathbf{\delta U} + \mathcal{L}(\mathbf{\delta C}), \label{eq:Ump}\\
\partial_t \mathbf{\delta C} &= \mathcal{A}(\mathbf{\delta U}) \mathbf{C}_e + \mathbf{C}_e \mathcal{A}^\dagger(\mathbf{\delta U}) \nonumber \\ & \quad + \mathcal{A}(\mathbf{U}_e) \mathbf{\delta C} + \mathbf{\delta C} \mathcal{A}^\dagger(\mathbf{U}_e), 
\label{eq:Cp}
\end{align}
\end{subequations}
subject to $\nabla \cdot \mathbf{\delta U} = 0$. Eigenanalysis of the linear operator in Eq. \eqref{eq:S3Tp} determines the S3T stability of the equilibrium state $(\mathbf{U}_e,\C_e)$. When the fluctuation variance that is controlled by $\Q$  exceeds a critical threshold the spanwise-homogeneous equilibrium becomes unstable with associated unstable eigenmodes of  RSS form. This S3T instability breaks the spanwise homogeneity
of the equilibrium state and grows exponentially, leading either to an equilibrium or a turbulent state depending on 
the excitation, $\Q$  \cite{Farrell-Ioannou-2012,Farrell-Ioannou-2017-bifur}.

\begin{figure}
    \centering
    \includegraphics[width=1.\linewidth]{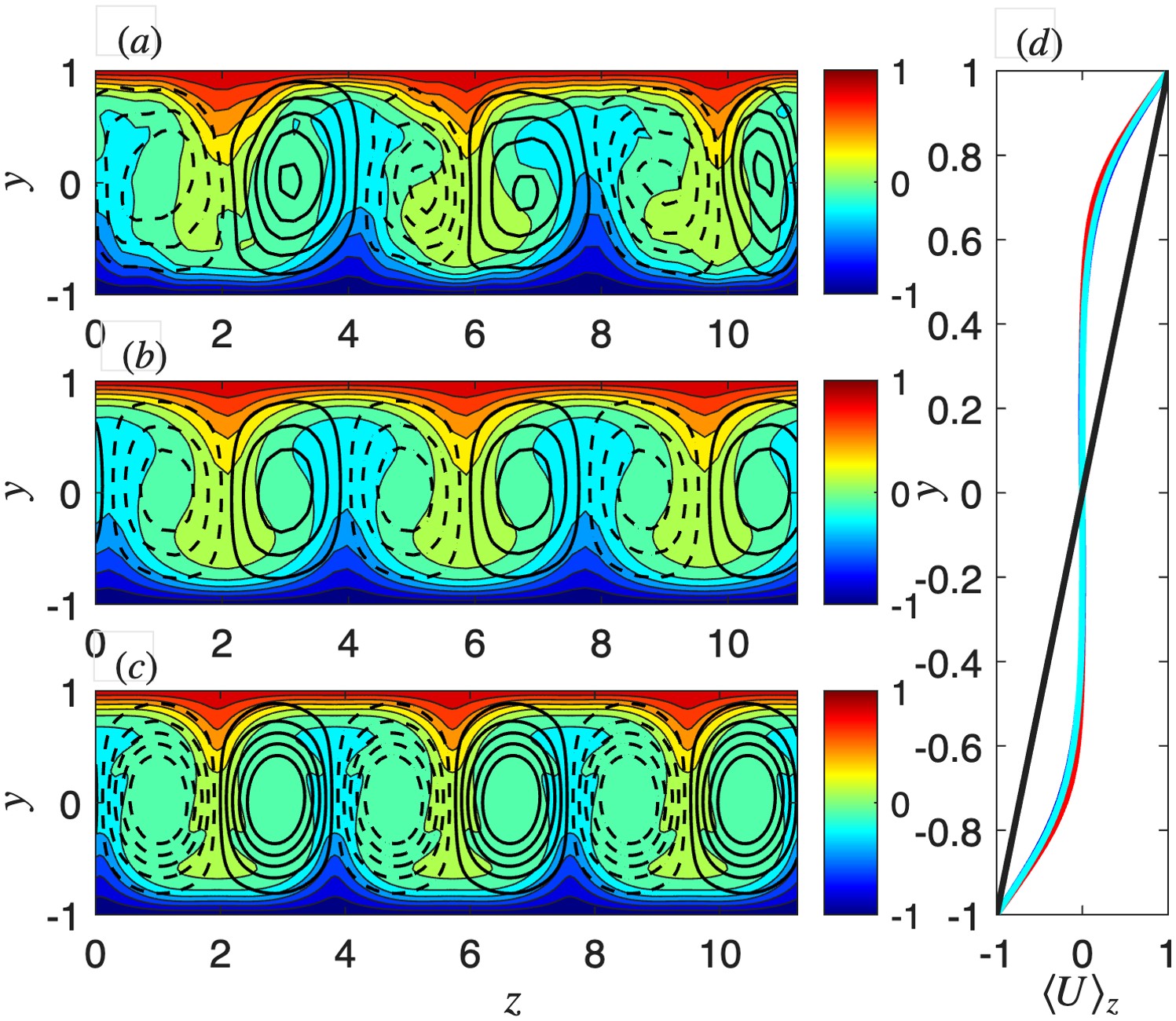}
    \caption{Mean-flow structure of the DNS and S3T turbulent states in a $L_z=3.6\pi$ channel at $R=600$. \textbf{(a)}--\textbf{(c)} $(y,z)$ cross-sections: 12 color contours of streamwise velocity $U$ with 10 equally spaced streamfunction contours $\Psi\in[-0.05,0.05]$, showing \textbf{(a)} an instantaneous DNS snapshot, \textbf{(b)} the DNS time mean ($\langle U\rangle_t$, $\langle\Psi\rangle_t$), and \textbf{(c)} the S3T equilibrium. \textbf{(d)} Spanwise-averaged profiles: instantaneous DNS $\langle U\rangle_z$ (cyan), DNS time-and-spanwise mean $\langle U\rangle_{z,t}$ (blue), S3T equilibrium $\langle U\rangle_z$ (red), and laminar Couette (black).}
    \label{fig:S3T_DNS}
   \vskip-0.2in
\end{figure}

\begin{figure}
    \centering
    \includegraphics[width=\linewidth]{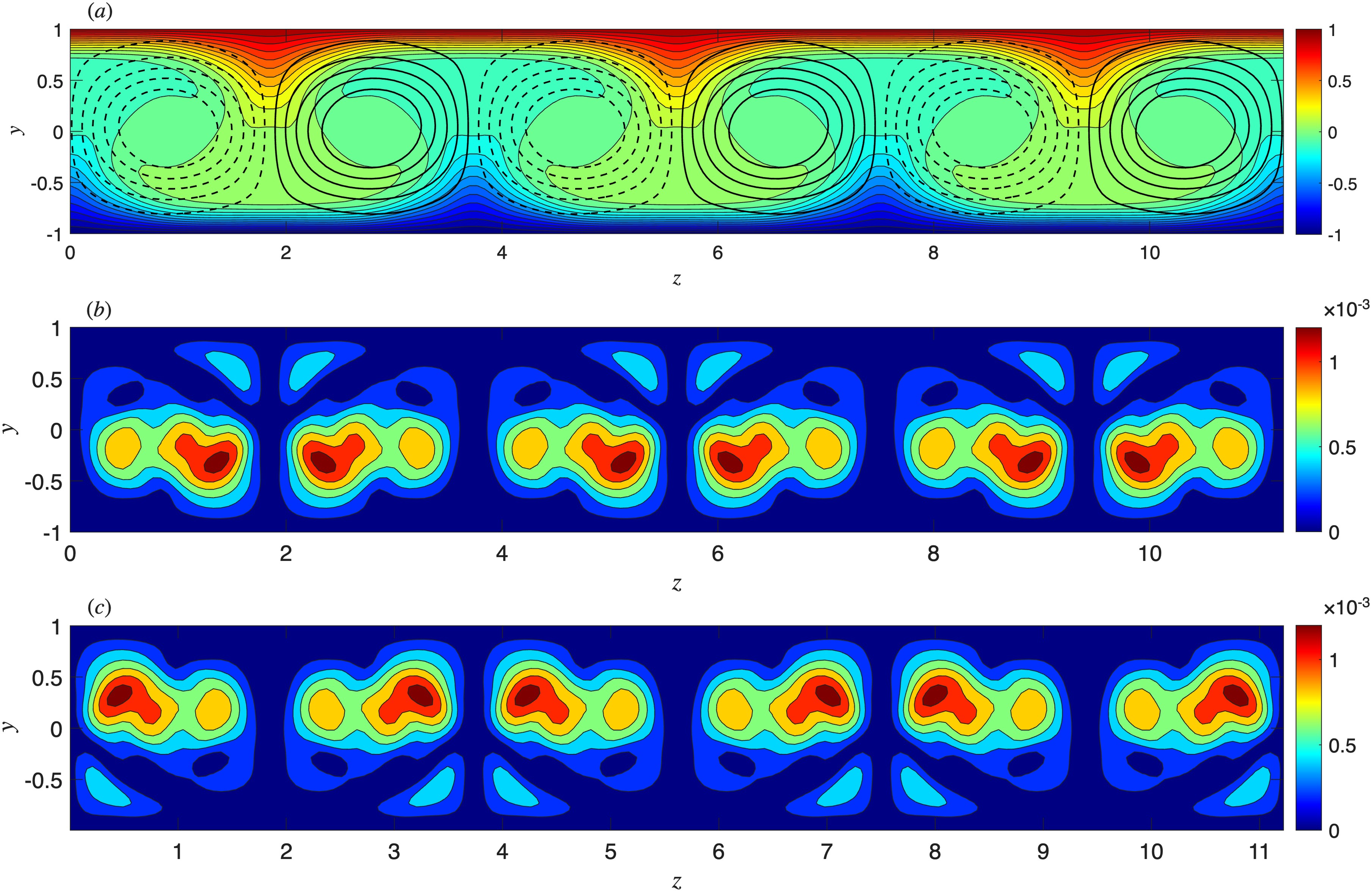}
    \caption{\footnotesize The three structures supporting the S3T stable equilibrium at $L_z=3.6\pi$, $R=600$. (a) First cumulant: streamwise velocity $U$ (color) and streamfunction $\Psi$ (contours). (b),(c) The two neutral modes of ${\cal A}(\mathbf{U})$ comprising the second cumulant, indicated by color contours of the fluctuation cross-stream velocity $|v'|$ in the $(y,z)$ plane. 
    The mode in (b), phase speed $c_r=0.12$, is associated with the high-speed streaks; that in (c), $c_r=-0.12$, with the low-speed streaks.}
    \label{fig:rank3}
    \vskip-0.2in
\end{figure}

We perform the S3T stability analysis for Couette flow in the presence of
a weak turbulent background in a channel with
$L_x \times L_z = 1.75\pi \times 3.6\pi$ at $R=600$. The fluctuation field
is maintained by stochastic excitation, white in time and homogeneous in
$x$ and $z$, parameterized by the covariance matrix $\Q$ (cf. \cite{Farrell-Ioannou-2012}). Because
$(\mathbf{U}_e,\C_e)$ is homogeneous in both $x$ and $z$, the
eigenfunctions are harmonic in these directions, with streamwise
wavenumber $k_x$ and spanwise mode number $m$. With $\Q$ chosen to
maintain fluctuation energy density $0.56\%$ of the laminar-flow energy
density at each streamwise wavenumber, the growth rate is maximized at
the gravest streamwise wavenumber, $k_x = 2\pi/L_x$; growth rates at this
$k_x$ as a function of $m$ are shown in Fig.~\ref{fig:S3Tstab}(a). The
most unstable eigenfunction has $m=6$ and its mean-flow component
$\delta\mathbf{U}$ is shown in Fig.~\ref{fig:S3Tstab}(b). When stochastic
excitation maintaining the same fluctuation energy density is introduced
into a DNS of plane Couette flow, the six-streak structure grows as
predicted (cf.~\cite{Farrell-Ioannou-2017-bifur,Nikolaidis-2024}).
Subsequently, the six RSSs undergo vortex merger and the flow transitions
to a turbulent state dominated by three RSSs. These structures eventually
settle into a spatially coherent quasi-equilibrium with minor temporal
fluctuations, as illustrated by the DNS snapshot in
Fig.~\ref{fig:S3T_DNS}(a). These three streaks remain remarkably coherent
[cf.~Fig.~\ref{fig:Hov}(a)], oscillating weakly around their time-mean
structure [Fig.~\ref{fig:S3T_DNS}(b)]. The S3T dynamics captures this
entire transition sequence and the resulting turbulent state
[Eqs.~\eqref{eq:S3Tm} and \eqref{eq:C}]. Following the initial linear
growth of the six-streak S3T instability and its subsequent vortex merger
and emergence of a three RSS structure, the S3T simulation reaches a
statistically stable state with slight temporal fluctuations that mirror
the DNS results [Figs.~\ref{fig:S3T_DNS}(c,d)]. Importantly, when the
external stochastic excitation is entirely removed ($\Q = 0$), the
fluctuation covariance at every streamwise harmonic other than the
gravest decays to zero, and the S3T state within this $L_z = 3.6\pi$
domain converges to the  fixed-point equilibrium state consisting of
the three RSSs shown in Fig.~\ref{fig:S3T_DNS}(c) and
Fig.~\ref{fig:rank3}: confinement of the equilibrium fluctuations to
$k_x = 2\pi/L_x$ is selected by the dynamics, not imposed. Because this
equilibrium is a $\Q=0$ fixed point, it is independent of the excitation
model: $\Q$ only seeds the linear instability that selects the basin of
attraction and plays no role in the attractor itself. The S3T mean flow
shown in Fig.~\ref{fig:S3T_DNS}(c) agrees with the DNS mean state
(for more details see~\cite{Comparison-Supplement}).

The stable equilibrium  statistical state in this $L_z = 3.6\pi$
domain is supported at the single gravest streamwise wavenumber $k_x$ by a rank 2 
covariance matrix $\mathbf{C}$. Together with the rank 1 mean state, this identifies 
the turbulent state in the S3T state space as comprising  three large-scale
analytically characterized structures: 
streaks and rolls of comparable amplitude and two eigenmodes of $\mathbf{C}$. Filtered LES 
simulations confirm that  RSSs in Couette turbulence similarly comprise rolls and streaks of comparable amplitude 
supported by large-scale fluctuations \cite{Rawat-etal-2015}.

The low rank of the fluctuation 
dynamics supporting the turbulent state follows directly from the structure of the 
Lyapunov equation [Eq.~\eqref{eq:C}] in the unforced limit $\mathbf{Q} = 0$ 
on noting that $\mathbf C$ must be composed  
of the neutral Lyapunov vectors of the linear operator $\mathcal{A}(\mathbf{U})$. 
Were any exponent positive the state would diverge; were the spectrum strictly stable it would laminarize. 
S3T turbulence exploits
Reynolds-stress feedback to modify the mean flow, driving the  originally unstable Lyapunov vectors to precisely zero exponent and holding them there.  These neutral modes of $\mathcal{A}(\mathbf{U})$, with phase speed $c_r=\pm 0.12$, shown in Fig. \ref{fig:rank3}, support the fixed point of S3T and identify the active subspace of the S3T  turbulent state.
In this way, Malkus's classic conjecture~\citep{Malkus-1956,*Reynolds-Tiederman-1967} 
is given an explicit realization within the S3T closure: turbulence is maintained by a neutral mean profile—with the proviso  that this mean state is the streamwise-mean and not the  planar averaged flow. Additionally, while Townsend~\citep{Townsend-1961}  posited the distinction between active and passive motions in shear turbulence, the S3T framework  supplies its
analytic characterization.

  In common with viscous critical layer modes underlying traditional SSP, VWI and ECS theory \cite{Nagata-1990,*Waleffe-1998,*Waleffe-2001,*Waleffe-2003,*Wang-etal-2007,*Viswanath-2007,*Gibson-etal-2009,*Hall-Sherwin-2010,*Blackburn-Hall-2013,*Deguchi-etal-2013,*Deguchi-Hall-2014}, these modes of $\mathcal{A}(\mathbf{U})$ are maintained by transfer of energy from the shear of the streamwise mean flow $U(y,z)$. However, it is crucial to note that the fluctuation eigenmodes underlying the SSP of Couette turbulence are not viscous critical layer eigenmodes.  In fact, scaling arguments show that the roll velocity in Couette turbulence is highly dominant over the diffusive term in the dynamics that regularizes the critical layer~
  ~\cite{Scaling-Supplement}.  While strong vortex RSS solutions have been previously studied 
  \cite{Jimenez-Kawahara-2005,*Kawahara-2009,*Kawahara-etal-2012,Rawat-etal-2015}, the role of these vortices in sustaining the
  neutral modes of the streamwise-mean flow was not identified.
  We find that these eigenmodes lie on a distinct manifold of instabilities  regularized by roll advection.  In these instabilities, the critical layer is replaced by a closed streamfunction of the roll circulation with Lagrangian circulation period resonant with that of the fluctuation eigenmode. 
This roll-advection regularized  Floquet mode (RRFM) is established and maintained by feedback regulation between the first and second cumulants.  This regulation, among other things, underlies the observed spanwise roll quantization. Analysis of RRFM manifold dynamics will be reported elsewhere.

\begin{figure}[h!]
    \centering
    \includegraphics[width=\columnwidth]{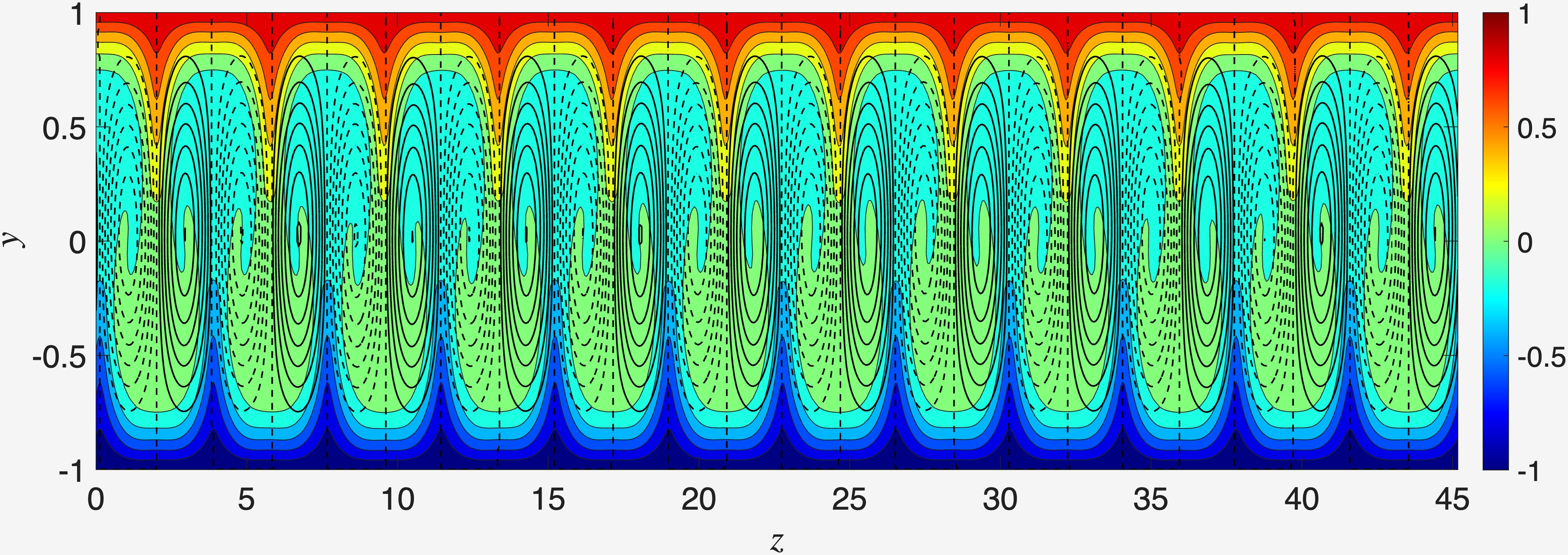}%{12HKWEQ55X720A.jpg}
    \caption{Time-mean streamwise velocity $\langle U\rangle_t$ (color) and roll streamfunction $\langle\Psi\rangle_t$ (black contours) of the S3T state in a $L_z=4\times3.6\pi$ channel at $R=600$. This wide-channel mean state is a 4-fold tiling of the spanwise wavenumber-3 equilibrium unit cell shown in Fig.~\ref{fig:S3T_DNS}(c).}
    \label{fig:HKW12}
    %\vskip-0.2in
\end{figure}

S3T equilibria exist for channels with spanwise widths near $L_z=3.6\pi$. We have also located an equilibrium near $L_z=2.2 \pi$ comprising a cell of two streaks~\cite{Farrell-Ioannou-2026}; however the three-streak unit forming the equilibrium at $L_z=3.6\pi$ is the robust and fundamental one that underlies the spanwise quantization. When the spanwise length is an integer multiple of this fundamental cell, the emerging S3T state is a tiling of the wavenumber-3 equilibrium shown in Fig.~\ref{fig:S3T_DNS}(c). For example, the S3T state at $L_z=4\times 3.6\pi$ and $R=600$ is shown in Fig.~\ref{fig:HKW12}.
Although any periodic tiling by the fundamental $L_z=3.6\pi$ equilibrium is itself an S3T equilibrium, all such tilings other than the fundamental are unstable to modulational instabilities. These instabilities equilibrate at small amplitude. For the fourfold tiling of Fig.~\ref{fig:HKW12}—a spanwise-wavenumber-12 state—the temporal fluctuations of the streamwise mean flow about the predicted mean have standard deviation order $2\%$.
 
In summary, the S3T SSD formulation of NS dynamics reveals a stable fixed point of Couette turbulence with spanwise wavenumber 3 that is analytically inaccessible to NS dynamics cast in traditional velocity variables.  
Among the results of this work are identifying a mechanism for understanding the long-observed coherence and spanwise quantization of WCC turbulence, revealing the large-scale organization of WCC  turbulence to be captured by its S3T rank 3 manifestation, analytically characterizing these structures, revealing these structures to comprise the active subspace supporting the turbulence, and demonstrating that the fluctuation component of Couette turbulence is supported on a manifold distinct from the traditionally assumed manifold of viscously regulated critical layer modes. 
Whether the same  manifold organizes the turbulence in pipe and channel flow is the natural next question.

\begin{acknowledgments}
We thank Professors Javier Jim\'enez and Myoungkyu Lee for insightful
discussions. Preliminary results of the SSD analysis of WCC turbulence were reported
in the conference proceedings~\cite{Farrell-Ioannou-2026} of the
2025 Madrid Turbulence Summer Workshop. 
This work was supported in part  by the European Research Council under the Caust grant
ERC-AdG-101018287.
\end{acknowledgments}
\appendix

\section{Numerical methods and integration parameters}
Both the DNS and the S3T integrations solve the dynamics in the form of
evolution equations for the wall-normal vorticity and the Laplacian of
the wall-normal velocity, as in Kim~et~al.~\citep{Kim-etal-1987}. Fields
are discretized by Fourier expansion in the streamwise and spanwise
directions, with dealiasing, and by Chebyshev collocation in the
wall-normal direction. Time stepping is by a third-order semi-implicit
Runge--Kutta method, with the viscous terms treated implicitly. All
integrations were performed with the vectorized ensemble direct
numerical simulation code of Nikolaidis~\citep{Nikolaidis-2024}.

For the channel with $L_x = 1.75\pi$ and $L_z = 3.6\pi$ at $R = 600$ we
used $N_z = 180$ equally spaced points in $z$ and $N_y = 53$ Chebyshev
collocation points in $y$; for channels with $L_z$ a multiple of
$3.6\pi$, $N_z$ was increased proportionally. The DNS retained 54
streamwise harmonics. For the S3T integrations it was first confirmed,
by integrations retaining the full set of streamwise wavenumbers, that
only the $k_x = 0$ and $k_x = 2\pi/L_x$ components are sustained when
$\Q = 0$; subsequent integrations therefore retained only these two
wavenumbers. Convergence of the S3T equilibria was verified by doubling
the number of discretization points. The stability of the fixed points
was determined by Arnoldi iteration on the linearized dynamics about the fixed streamwise-mean flow,
using the same code as the time stepper.

\section{Comparison between the S3T stable equilibrium and the corresponding DNS turbulent state}

\begin{figure}
    \centering
    \includegraphics[width=1\linewidth]{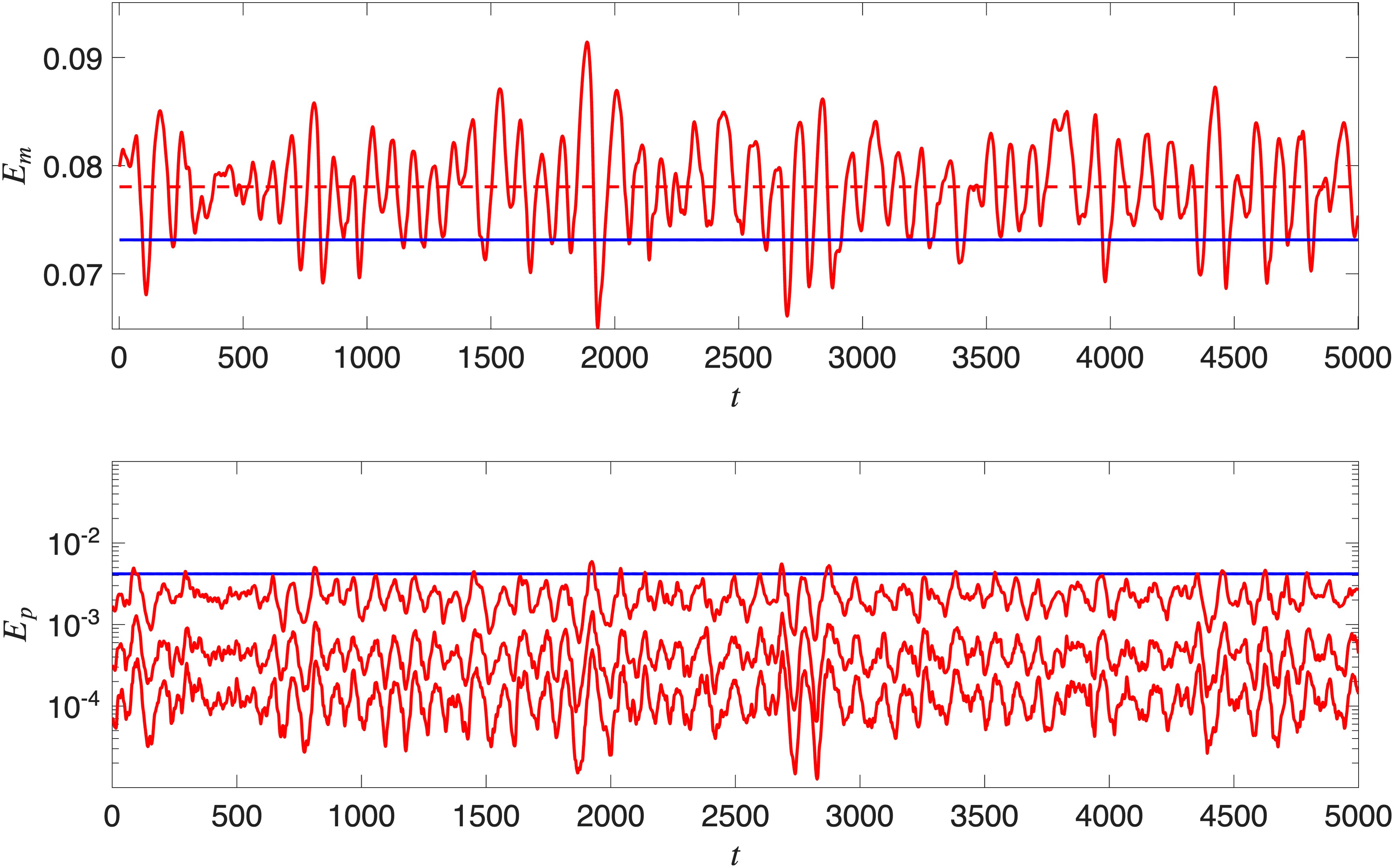}
    \caption{Top panel:  Kinetic energy density of the streamwise mean flow, $E_m = \tfrac12|\Uv|^2$.  The $E_m$ of the DNS is shown in red and the dashed red line is its time mean.
    % , $\tfrac12\langle|\Uv|^2\rangle_t$.
The $E_m$ of the streamwise mean flow of the S3T equilibrium (c.f. Fig. 3(c) of the Letter) is shown in blue. 
Bottom panel: Kinetic energy, $E_p$, of the streamwise-varying fluctuations in the DNS 
(red), resolved by streamwise wavenumber. The energy decreases monotonically with $k_x$: most of it resides in the 
gravest harmonic, $k_x = 2\pi/L_x$ (uppermost red curve), followed by the second and third harmonics. The solid blue line is 
the kinetic energy density of the two neutral modes that support the S3T equilibrium. 
For $L_x \times L_z = 1.75\pi \times 3.6\pi$ at $R = 600$.}
    \label{fig:E0}
\end{figure}

Kinetic energy density of the streamwise mean flow and of the dominant streamwise Fourier components of the fluctuations in the DNS  together with the corresponding Fourier components of the S3T equilibrium for Couette turbulence in channel $L_z = 3.6\pi$  at $R = 600$ are shown in Figure~\ref{fig:E0}. The S3T equilibrium supports fluctuations only at the streamwise wavenumber $k_x = 2\pi/L_x$,
indicating that, for this channel geometry and Reynolds number, the active subspace sustaining the turbulence consists of this single harmonic. 
Consistent with this, the DNS energy distribution in Fig.~\ref{fig:E0} (bottom panel) is dominated by the $k_x = 2\pi/L_x$ component.  
As shown in Figs.~1 and 3 of the Letter, the streamwise 
mean flow in the DNS exhibits small temporal fluctuations 
about its time mean, with  normalized RMS amplitude $5\%$, and remains spatially coherent at all times.
However, temporal fluctuations of the streamwise varying components are substantial: the normalized RMS amplitude of fluctuations of the dominant streamwise wavenumber in the DNS
$k_x= 2 \pi/L_x$ is $37 \%$. This indicates that the variability of the fluctuation field does not strongly influence the streamwise-mean flow: the mean Reynolds-stress forcing by the fluctuations remains robust when averaged over the long time scale of the mean RSS. The fluctuation components that sustain the streamwise mean flow retain their RRFM form (defined in the Letter)
in DNS but become time-dependent.   Specifically, the RRFM associated with individual RSS is independently varying in DNS so that while the time mean stresses are nearly identical in S3T and DNS, the global synchronization of the fluctuation component in S3T is not preserved in DNS.

% in DNS but become time-dependent: the forcings associated with individual RSS vary independently, so that although the time-mean stresses are nearly identical in S3T and DNS, the global synchronization of the fluctuation components that holds in the S3T equilibrium is not maintained in DNS.

% {\color{red} The  fluctuation components that sustain  the streamwise mean flow in S3T retain their RRFM form in DNS but become time-dependent.   Specifically, the RRFM associated with individual RSS are independently varying in DNS so that while the time mean stresses are nearly identical in S3T and DNS, the global synchronization of the fluctuation component in S3T is not preserved in DNS.}

The most noteworthy difference between the S3T equilibrium and the DNS in their respective means is the strength 
of the roll circulation. The maximum roll velocities in the S3T state reach $0.11$,
nearly double the corresponding DNS values of $0.06$ (cf. Fig.~3 of the Letter). 
Remarkably, despite this difference 
the turbulent mean profiles of the S3T state and the DNS are almost identical 
(cf. Fig.~3d of the Letter), and consistently the two flows share essentially the same friction Reynolds number: $R_\tau=46$
for the S3T fixed-point state,  $R_\tau=44$ for the time mean of the DNS. Equal $R_\tau$
implies equal wall stress and hence a nearly identical total rate of energy dissipation.
This near-equality of $R_\tau$
predicted by S3T and observed in fully nonlinear simulations has previously been noted to persist at high Reynolds number \cite{Bretheim-etal-2015,*Farrell-etal-2016-VLSM,*Bretheim-etal-2018}.
We note that the $V$ and $W$ in DNS are still large enough to
place the flow  in the regime in which
fluctuation regularization occurs by roll advection rather than by viscosity.

\section{Scaling relation comparing the relevance of viscosity and advection  in regularizing neutral mode structure at a critical layer in shear flow}

The linear evolution of velocity perturbations $\uv'$ to a streamwise mean
flow $\Uv=(U,V,W)$ is governed by
\begin{gather}
\partial_t \uv' + (\Uv \cdot \nablav) \uv' - \frac{1}{R} \Delta \uv'
= \Fv~, \qquad \nablav \cdot \uv' = 0~, \label{eq:A}
\end{gather}
where the production term is
\begin{gather*}
\Fv \equiv - (\uv' \cdot \nablav) \Uv - \nablav p'~.
\end{gather*}
Assuming perturbation mode form $\uv' = \hat{\uv}\, e^{ik_x(x-ct)}$, Eq.~\eqref{eq:A}
becomes
\begin{gather}
\left( ik_x(U(y,z)-c) + V(y,z)\partial_y + W(y,z)\partial_z
- \frac{1}{R}\Delta \right) \hat{\uv} = \hat{\Fv}~, \label{eq:A1}
\end{gather}
with $\hat{\Fv}$ the Fourier amplitude of $\Fv$.

In the case of a plane-parallel flow $\Uv(y,z) = (U(y,z),0,0)$ Eq.~\eqref{eq:A1} admits a 
neutral eigenmode with phase speed $c_r$ in which the structure of the solution in the vicinity of the critical layer $U(y,z)=c$ 
is determined by the viscous term. The requirement for this viscous regularization is
that within a region of width $\delta_v$ about the critical layer,
Eq.~\eqref{eq:A1} reduces to the balance
\begin{gather}
ik_x(U(y,z)-c)\hat{\uv} \approx \frac{1}{R}\Delta\hat{\uv}~. \label{eq:A1v}
\end{gather}
The width $\delta_v$ of this region then satisfies
\begin{gather*}
k_x U_c' \delta_v \approx \frac{1}{R\delta_v^2}~,
\end{gather*}
where $U_c' \equiv \|\nabla U\|_{U(y,z)=c}$ is the magnitude of the mean shear
in the direction normal to the critical layer. Hence viscous
regularization~\cite{Drazin-Reid-81,*Schmid-Henningson-2001} dominates in a region of width
\begin{gather}
\delta_v = (k_xRU_c')^{-1/3}~. \label{eq:deltav}
\end{gather}

However, in the presence of a mean streamwise roll, i.e.\ with $V$ and $W$ not vanishing, the behavior of the solution is
determined  by the advection term with the alternative balance,
\begin{gather*}
ik_x(U(y,z)-c)\hat{\uv} \approx  -\left(V(y,z)\partial_y + W(y,z)\partial_z\right)\hat{\uv}~,
\end{gather*}
which extends over a region of thickness $\delta_r$ satisfying
\begin{gather*}
k_x U_c' \delta_r \approx \frac{V_n}{\delta_r}~,
\end{gather*}
where $V_n$ is the magnitude of  $(V,W)$ in the direction
normal to the critical layer.  It follows that the influence of  roll advection on the structure of the neutral mode
extends over a region of width:
\begin{gather}
\delta_r = \left(\frac{V_n}{k_xU_c'}\right)^{1/2}~. \label{eq:delta_r}
\end{gather}

The ratio of these widths  measures the relative importance of viscosity and advection 
in determining the structure of the modes that sustain the SSP:  
\begin{gather}
\frac{\delta_r}{\delta_v}=(k_xRU_c')^{1/3}\left(\frac{V_n}{k_xU_c'}\right)^{1/2}
=R^{1/3}\,V_n^{1/2}\,(k_xU_c')^{-1/6} \label{eq:deltavoverdelta_r}
\end{gather}

The stable S3T equilibrium at $L_z=3.6\pi$ and $R=600$ shown in Fig.~3 has roll
velocities of $O(10^{-1})$. Even at this moderate Reynolds number, the
$\delta_r$ associated with the neutral eigenmodes supporting the equilibrium is
significantly larger than $\delta_v$. With $k_x\approx 1.14$ at $R=600$ and $U_c'\approx 1$  we obtain $\delta_v \approx 0.11$. For the S3T
with  $V_n\approx 0.11$ we obtain $\delta_r \approx 0.31 $  and for the DNS with $V_n\approx 0.06$ we obtain  $\delta_r \approx 0.23 $, with both of these widths comparable in size to the half-channel.
This is consistent with  the neutral modes supporting the equilibrium turbulent state in WCC being roll-advection
dominated and suggests existence of a manifold of eigenmodes distinct from  the viscously
regularized modes that support the lower-branch equilibria in VWI
theory~\cite{Hall-Sherwin-2010,*Blackburn-Hall-2013,%
*Deguchi-etal-2013,*Deguchi-Hall-2014} and the corresponding lower-branch
ECS~\cite{Nagata-1990,*Waleffe-2001,*Waleffe-2003,*Wang-etal-2007} which have $O(1/R)$
roll velocities and are also viscously regularized. 

That roll-advection domination of the mode  structure persists  at high Reynolds numbers is implied by the LES results
of Cheng et al. (2022) \cite{Cheng-Pullin-2022} (cf. their Figures 19b, 25d, 26 and Equation (7.7))
showing that up to $R\sim 10^5$ the roll velocity
decays at the rate $\approx 1/\log R$. This is far
slower than the $V_n\sim R^{-2/3}$ decay (the crossover obtained by equating $\delta_r=\delta_v$) required for
asymptotic dominance of viscous regularization, so roll-advection regularization becomes increasingly dominant
as $R$ increases. This is consistent with 
the persistence of the WCC  near equilibrium structure at high Reynolds number seen in  DNS and LES \cite{Pirozzoli-etal-2014, *Avsarkisov-etal-2014,*Rawat-etal-2015,*Lee-Moser-2018,*Cheng-Pullin-2022, *Hoyas-Oberlack-2024}.

\bibliography{basic_references}

%apsrev4-2.bst 2019-01-14 (MD) hand-edited version of apsrev4-1.bst
%Control: key (0)
%Control: author (8) initials jnrlst
%Control: editor formatted (1) identically to author
%Control: production of article title (0) allowed
%Control: page (0) single
%Control: year (1) truncated
%Control: production of eprint (0) enabled
\begin{thebibliography}{55}%
\makeatletter
\providecommand \@ifxundefined [1]{%
 \@ifx{#1\undefined}
}%
\providecommand \@ifnum [1]{%
 \ifnum #1\expandafter \@firstoftwo
 \else \expandafter \@secondoftwo
 \fi
}%
\providecommand \@ifx [1]{%
 \ifx #1\expandafter \@firstoftwo
 \else \expandafter \@secondoftwo
 \fi
}%
\providecommand \natexlab [1]{#1}%
\providecommand \enquote  [1]{``#1''}%
\providecommand \bibnamefont  [1]{#1}%
\providecommand \bibfnamefont [1]{#1}%
\providecommand \citenamefont [1]{#1}%
\providecommand \href@noop [0]{\@secondoftwo}%
\providecommand \href [0]{\begingroup \@sanitize@url \@href}%
\providecommand \@href[1]{\@@startlink{#1}\@@href}%
\providecommand \@@href[1]{\endgroup#1\@@endlink}%
\providecommand \@sanitize@url [0]{\catcode `\\12\catcode `\$12\catcode
  `\&12\catcode `\#12\catcode `\^12\catcode `\_12\catcode `\%12\relax}%
\providecommand \@@startlink[1]{}%
\providecommand \@@endlink[0]{}%
\providecommand \url  [0]{\begingroup\@sanitize@url \@url }%
\providecommand \@url [1]{\endgroup\@href {#1}{\urlprefix }}%
\providecommand \urlprefix  [0]{URL }%
\providecommand \Eprint [0]{\href }%
\providecommand \doibase [0]{https://doi.org/}%
\providecommand \selectlanguage [0]{\@gobble}%
\providecommand \bibinfo  [0]{\@secondoftwo}%
\providecommand \bibfield  [0]{\@secondoftwo}%
\providecommand \translation [1]{[#1]}%
\providecommand \BibitemOpen [0]{}%
\providecommand \bibitemStop [0]{}%
\providecommand \bibitemNoStop [0]{.\EOS\space}%
\providecommand \EOS [0]{\spacefactor3000\relax}%
\providecommand \BibitemShut  [1]{\csname bibitem#1\endcsname}%
\let\auto@bib@innerbib\@empty
%</preamble>
\bibitem [{\citenamefont {Bech}\ and\ \citenamefont
  {Andersson}(1994)}]{Bech-1994}%
  \BibitemOpen
  \bibfield  {author} {\bibinfo {author} {\bibfnamefont {K.~H.}\ \bibnamefont
  {Bech}}\ and\ \bibinfo {author} {\bibfnamefont {H.~I.}\ \bibnamefont
  {Andersson}},\ }\bibinfo {title} {Very-large-scale structures in {DNS}},\ in\
  \href {https://doi.org/10.1007/978-94-011-1000-6_2} {\emph {\bibinfo
  {booktitle} {Direct and Large-Eddy Simulation I: Selected papers from the
  First ERCOFTAC Workshop on Direct and Large-Eddy Simulation}}},\ \bibinfo
  {editor} {edited by\ \bibinfo {editor} {\bibfnamefont {P.~R.}\ \bibnamefont
  {Voke}}, \bibinfo {editor} {\bibfnamefont {L.}~\bibnamefont {Kleiser}},\ and\
  \bibinfo {editor} {\bibfnamefont {J.-P.}\ \bibnamefont {Chollet}}}\ (\bibinfo
   {publisher} {Springer Netherlands},\ \bibinfo {address} {Dordrecht},\
  \bibinfo {year} {1994})\ pp.\ \bibinfo {pages} {13--24}\BibitemShut {NoStop}%
\bibitem [{\citenamefont {Papavassiliou}\ and\ \citenamefont
  {Hanratty}(1997)}]{Papavasiliou-1997}%
  \BibitemOpen
  \bibfield  {author} {\bibinfo {author} {\bibfnamefont {D.~V.}\ \bibnamefont
  {Papavassiliou}}\ and\ \bibinfo {author} {\bibfnamefont {T.~J.}\ \bibnamefont
  {Hanratty}},\ }\bibfield  {title} {\bibinfo {title} {Interpretation of
  large-scale structures observed in a turbulent plane {Couette} flow},\ }\href
  {https://doi.org/https://doi.org/10.1016/S0142-727X(96)00138-5} {\bibfield
  {journal} {\bibinfo  {journal} {International Journal of Heat and Fluid
  Flow}\ }\textbf {\bibinfo {volume} {18}},\ \bibinfo {pages} {55} (\bibinfo
  {year} {1997})}\BibitemShut {NoStop}%
\bibitem [{\citenamefont {Tillmark}\ and\ \citenamefont
  {Alfredsson}(1995)}]{Tillmark-1995}%
  \BibitemOpen
  \bibfield  {author} {\bibinfo {author} {\bibfnamefont {N.}~\bibnamefont
  {Tillmark}}\ and\ \bibinfo {author} {\bibfnamefont {P.~H.}\ \bibnamefont
  {Alfredsson}},\ }\bibfield  {title} {\bibinfo {title} {Structures in
  turbulent plane {Couette} flow obtained from correlation measurements},\ }in\
  \href@noop {} {\emph {\bibinfo {booktitle} {Advances in Turbulence V}}},\
  \bibinfo {editor} {edited by\ \bibinfo {editor} {\bibfnamefont
  {R.}~\bibnamefont {Benzi}}}\ (\bibinfo  {publisher} {Springer Netherlands},\
  \bibinfo {address} {Dordrecht},\ \bibinfo {year} {1995})\ pp.\ \bibinfo
  {pages} {502--507}\BibitemShut {NoStop}%
\bibitem [{\citenamefont {Tillmark}\ and\ \citenamefont
  {Alfredsson}(1998)}]{Tillmark-1998}%
  \BibitemOpen
  \bibfield  {author} {\bibinfo {author} {\bibfnamefont {N.}~\bibnamefont
  {Tillmark}}\ and\ \bibinfo {author} {\bibfnamefont {P.~H.}\ \bibnamefont
  {Alfredsson}},\ }\bibfield  {title} {\bibinfo {title} {Large scale structures
  in turbulent plane {Couette} flow},\ }in\ \href@noop {} {\emph {\bibinfo
  {booktitle} {Advances in Turbulence VII}}},\ \bibinfo {editor} {edited by\
  \bibinfo {editor} {\bibfnamefont {U.}~\bibnamefont {Frisch}}}\ (\bibinfo
  {publisher} {Springer Netherlands},\ \bibinfo {address} {Dordrecht},\
  \bibinfo {year} {1998})\ pp.\ \bibinfo {pages} {59--62}\BibitemShut {NoStop}%
\bibitem [{\citenamefont {Kitoh}\ \emph {et~al.}(2005)\citenamefont {Kitoh},
  \citenamefont {Nakabayashi},\ and\ \citenamefont {Nishimura}}]{Kitoh-2005}%
  \BibitemOpen
  \bibfield  {author} {\bibinfo {author} {\bibfnamefont {O.}~\bibnamefont
  {Kitoh}}, \bibinfo {author} {\bibfnamefont {K.}~\bibnamefont {Nakabayashi}},\
  and\ \bibinfo {author} {\bibfnamefont {F.}~\bibnamefont {Nishimura}},\
  }\bibfield  {title} {\bibinfo {title} {Experimental study on mean velocity
  and turbulence characteristics of plane {Couette} flow: low-{Reynolds}-number
  effects and large longitudinal vortical structure},\ }\href
  {https://doi.org/10.1017/S0022112005005641} {\bibfield  {journal} {\bibinfo
  {journal} {Journal of Fluid Mechanics}\ }\textbf {\bibinfo {volume} {539}},\
  \bibinfo {pages} {199–227} (\bibinfo {year} {2005})}\BibitemShut {NoStop}%
\bibitem [{\citenamefont {Kitoh}\ and\ \citenamefont
  {Umeki}(2008)}]{Kitoh-2008}%
  \BibitemOpen
  \bibfield  {author} {\bibinfo {author} {\bibfnamefont {O.}~\bibnamefont
  {Kitoh}}\ and\ \bibinfo {author} {\bibfnamefont {M.}~\bibnamefont {Umeki}},\
  }\bibfield  {title} {\bibinfo {title} {Experimental study on large-scale
  streak structure in the core region of turbulent plane {Couette} flow},\
  }\href {https://doi.org/10.1063/1.2844476} {\bibfield  {journal} {\bibinfo
  {journal} {Physics of Fluids}\ }\textbf {\bibinfo {volume} {20}},\ \bibinfo
  {pages} {025107} (\bibinfo {year} {2008})}\BibitemShut {NoStop}%
\bibitem [{\citenamefont {Komminaho}\ \emph {et~al.}(1996)\citenamefont
  {Komminaho}, \citenamefont {Lundbladh},\ and\ \citenamefont
  {Johansson}}]{Komminaho-etal-1996}%
  \BibitemOpen
  \bibfield  {author} {\bibinfo {author} {\bibfnamefont {J.}~\bibnamefont
  {Komminaho}}, \bibinfo {author} {\bibfnamefont {A.}~\bibnamefont
  {Lundbladh}},\ and\ \bibinfo {author} {\bibfnamefont {A.}~\bibnamefont
  {Johansson}},\ }\bibfield  {title} {\bibinfo {title} {Very large structures
  in plane turbulent {Couette} flow},\ }\href
  {https://doi.org/10.1017/S0022112096007537} {\bibfield  {journal} {\bibinfo
  {journal} {J. Fluid Mech.}\ }\textbf {\bibinfo {volume} {320}},\ \bibinfo
  {pages} {259} (\bibinfo {year} {1996})}\BibitemShut {NoStop}%
\bibitem [{\citenamefont {Tsukahara}\ \emph {et~al.}(2006)\citenamefont
  {Tsukahara}, \citenamefont {Kawamura},\ and\ \citenamefont
  {Shingai}}]{Tsukahara-etal-2006}%
  \BibitemOpen
  \bibfield  {author} {\bibinfo {author} {\bibfnamefont {T.}~\bibnamefont
  {Tsukahara}}, \bibinfo {author} {\bibfnamefont {H.}~\bibnamefont
  {Kawamura}},\ and\ \bibinfo {author} {\bibfnamefont {K.}~\bibnamefont
  {Shingai}},\ }\bibfield  {title} {\bibinfo {title} {{DNS} of turbulent
  {C}ouette flow with emphasis on the large-scale structure in the core
  region},\ }\href {https://doi.org/10.1080/14685240600609866} {\bibfield
  {journal} {\bibinfo  {journal} {Journal of Turbulence}\ }\textbf {\bibinfo
  {volume} {7}},\ \bibinfo {pages} {N19} (\bibinfo {year} {2006})}\BibitemShut
  {NoStop}%
\bibitem [{\citenamefont {Pirozzoli}\ \emph {et~al.}(2014)\citenamefont
  {Pirozzoli}, \citenamefont {Bernardini},\ and\ \citenamefont
  {Orlandi}}]{Pirozzoli-etal-2014}%
  \BibitemOpen
  \bibfield  {author} {\bibinfo {author} {\bibfnamefont {S.}~\bibnamefont
  {Pirozzoli}}, \bibinfo {author} {\bibfnamefont {M.}~\bibnamefont
  {Bernardini}},\ and\ \bibinfo {author} {\bibfnamefont {P.}~\bibnamefont
  {Orlandi}},\ }\bibfield  {title} {\bibinfo {title} {Turbulence statistics in
  {Couette} flow at high {Reynolds} number},\ }\href
  {https://doi.org/10.1017/jfm.2014.529} {\bibfield  {journal} {\bibinfo
  {journal} {Journal of Fluid Mechanics}\ }\textbf {\bibinfo {volume} {758}},\
  \bibinfo {pages} {327} (\bibinfo {year} {2014})}\BibitemShut {NoStop}%
\bibitem [{\citenamefont {Avsarkisov}\ \emph {et~al.}(2014)\citenamefont
  {Avsarkisov}, \citenamefont {Hoyas}, \citenamefont {Oberlack},\ and\
  \citenamefont {Garcia-Galache}}]{Avsarkisov-etal-2014}%
  \BibitemOpen
  \bibfield  {author} {\bibinfo {author} {\bibfnamefont {V.}~\bibnamefont
  {Avsarkisov}}, \bibinfo {author} {\bibfnamefont {S.}~\bibnamefont {Hoyas}},
  \bibinfo {author} {\bibfnamefont {M.}~\bibnamefont {Oberlack}},\ and\
  \bibinfo {author} {\bibfnamefont {J.~P.}\ \bibnamefont {Garcia-Galache}},\
  }\bibfield  {title} {\bibinfo {title} {Turbulent plane {Couette} flow at
  moderately high {Reynolds} number},\ }\href
  {https://doi.org/10.1017/jfm.2014.323} {\bibfield  {journal} {\bibinfo
  {journal} {J. Fluid Mech.}\ }\textbf {\bibinfo {volume} {751}},\ \bibinfo
  {pages} {R1} (\bibinfo {year} {2014})}\BibitemShut {NoStop}%
\bibitem [{\citenamefont {Rawat}\ \emph {et~al.}(2015)\citenamefont {Rawat},
  \citenamefont {Cossu}, \citenamefont {Hwang},\ and\ \citenamefont
  {Rincon}}]{Rawat-etal-2015}%
  \BibitemOpen
  \bibfield  {author} {\bibinfo {author} {\bibfnamefont {S.}~\bibnamefont
  {Rawat}}, \bibinfo {author} {\bibfnamefont {C.}~\bibnamefont {Cossu}},
  \bibinfo {author} {\bibfnamefont {Y.}~\bibnamefont {Hwang}},\ and\ \bibinfo
  {author} {\bibfnamefont {F.}~\bibnamefont {Rincon}},\ }\bibfield  {title}
  {\bibinfo {title} {On the self-sustained nature of large-scale motions in
  turbulent {Couette} flow},\ }\href {https://doi.org/10.1017/jfm.2015.550}
  {\bibfield  {journal} {\bibinfo  {journal} {J. Fluid Mech.}\ }\textbf
  {\bibinfo {volume} {782}},\ \bibinfo {pages} {515} (\bibinfo {year}
  {2015})}\BibitemShut {NoStop}%
\bibitem [{\citenamefont {Lee}\ and\ \citenamefont
  {Moser}(2018)}]{Lee-Moser-2018}%
  \BibitemOpen
  \bibfield  {author} {\bibinfo {author} {\bibfnamefont {M.}~\bibnamefont
  {Lee}}\ and\ \bibinfo {author} {\bibfnamefont {R.~D.}\ \bibnamefont
  {Moser}},\ }\bibfield  {title} {\bibinfo {title} {Extreme-scale motions in
  turbulent plane {Couette} flows},\ }\href
  {https://doi.org/10.1017/jfm.2018.131} {\bibfield  {journal} {\bibinfo
  {journal} {Journal of Fluid Mechanics}\ }\textbf {\bibinfo {volume} {842}},\
  \bibinfo {pages} {128} (\bibinfo {year} {2018})}\BibitemShut {NoStop}%
\bibitem [{\citenamefont {Cheng}\ \emph {et~al.}(2022)\citenamefont {Cheng},
  \citenamefont {Pullin},\ and\ \citenamefont {Samtaney}}]{Cheng-Pullin-2022}%
  \BibitemOpen
  \bibfield  {author} {\bibinfo {author} {\bibfnamefont {W.}~\bibnamefont
  {Cheng}}, \bibinfo {author} {\bibfnamefont {D.}~\bibnamefont {Pullin}},\ and\
  \bibinfo {author} {\bibfnamefont {R.}~\bibnamefont {Samtaney}},\ }\bibfield
  {title} {\bibinfo {title} {Wall-resolved and wall-modelled large-eddy
  simulation of plane {Couette} flow},\ }\href
  {https://doi.org/10.1017/jfm.2021.1046} {\bibfield  {journal} {\bibinfo
  {journal} {Journal of Fluid Mechanics}\ }\textbf {\bibinfo {volume} {934}},\
  \bibinfo {pages} {A19} (\bibinfo {year} {2022})}\BibitemShut {NoStop}%
\bibitem [{\citenamefont {Hoyas}\ and\ \citenamefont
  {Oberlack}(2024)}]{Hoyas-Oberlack-2024}%
  \BibitemOpen
  \bibfield  {author} {\bibinfo {author} {\bibfnamefont {S.}~\bibnamefont
  {Hoyas}}\ and\ \bibinfo {author} {\bibfnamefont {M.}~\bibnamefont
  {Oberlack}},\ }\bibfield  {title} {\bibinfo {title} {Turbulent {Couette} flow
  up to ${{Re}}_\tau =2000$},\ }\href {https://doi.org/10.1017/jfm.2024.369}
  {\bibfield  {journal} {\bibinfo  {journal} {Journal of Fluid Mechanics}\
  }\textbf {\bibinfo {volume} {987}},\ \bibinfo {pages} {R9} (\bibinfo {year}
  {2024})}\BibitemShut {NoStop}%
\bibitem [{\citenamefont {Farrell}\ and\ \citenamefont
  {Ioannou}(2003)}]{Farrell-Ioannou-2003-structural}%
  \BibitemOpen
  \bibfield  {author} {\bibinfo {author} {\bibfnamefont {B.~F.}\ \bibnamefont
  {Farrell}}\ and\ \bibinfo {author} {\bibfnamefont {P.~J.}\ \bibnamefont
  {Ioannou}},\ }\bibfield  {title} {\bibinfo {title} {Structural stability of
  turbulent jets},\ }\href
  {https://doi.org/10.1175/1520-0469(2003)060<2101:SSOTJ>2.0.CO;2} {\bibfield
  {journal} {\bibinfo  {journal} {J. Atmos. Sci.}\ }\textbf {\bibinfo {volume}
  {60}},\ \bibinfo {pages} {2101} (\bibinfo {year} {2003})}\BibitemShut
  {NoStop}%
\bibitem [{\citenamefont {Farrell}\ and\ \citenamefont
  {Ioannou}(2012)}]{Farrell-Ioannou-2012}%
  \BibitemOpen
  \bibfield  {author} {\bibinfo {author} {\bibfnamefont {B.~F.}\ \bibnamefont
  {Farrell}}\ and\ \bibinfo {author} {\bibfnamefont {P.~J.}\ \bibnamefont
  {Ioannou}},\ }\bibfield  {title} {\bibinfo {title} {Dynamics of streamwise
  rolls and streaks in turbulent wall-bounded shear flow},\ }\href
  {https://doi.org/10.1017/jfm.2012.300} {\bibfield  {journal} {\bibinfo
  {journal} {J. Fluid Mech.}\ }\textbf {\bibinfo {volume} {708}},\ \bibinfo
  {pages} {149} (\bibinfo {year} {2012})}\BibitemShut {NoStop}%
\bibitem [{\citenamefont {Farrell}\ and\ \citenamefont
  {Ioannou}(2007)}]{Farrell-Ioannou-2007-structure}%
  \BibitemOpen
  \bibfield  {author} {\bibinfo {author} {\bibfnamefont {B.~F.}\ \bibnamefont
  {Farrell}}\ and\ \bibinfo {author} {\bibfnamefont {P.~J.}\ \bibnamefont
  {Ioannou}},\ }\bibfield  {title} {\bibinfo {title} {Structure and spacing of
  jets in barotropic turbulence},\ }\href {https://doi.org/10.1175/JAS4016.1}
  {\bibfield  {journal} {\bibinfo  {journal} {J. Atmos. Sci.}\ }\textbf
  {\bibinfo {volume} {64}},\ \bibinfo {pages} {3652} (\bibinfo {year}
  {2007})}\BibitemShut {NoStop}%
\bibitem [{\citenamefont {Farrell}\ and\ \citenamefont
  {Ioannou}(2009{\natexlab{a}})}]{Farrell-Ioannou-2009-equatorial}%
  \BibitemOpen
  \bibfield  {author} {\bibinfo {author} {\bibfnamefont {B.~F.}\ \bibnamefont
  {Farrell}}\ and\ \bibinfo {author} {\bibfnamefont {P.~J.}\ \bibnamefont
  {Ioannou}},\ }\bibfield  {title} {\bibinfo {title} {Emergence of jets from
  turbulence in the shallow-water equations on an equatorial beta plane},\
  }\href {https://doi.org/10.1175/2009JAS2941.1} {\bibfield  {journal}
  {\bibinfo  {journal} {J. Atmos. Sci.}\ }\textbf {\bibinfo {volume} {66}},\
  \bibinfo {pages} {3197} (\bibinfo {year} {2009}{\natexlab{a}})}\BibitemShut
  {NoStop}%
\bibitem [{\citenamefont {Farrell}\ and\ \citenamefont
  {Ioannou}(2009{\natexlab{b}})}]{Farrell-Ioannou-2009-plasmas}%
  \BibitemOpen
  \bibfield  {author} {\bibinfo {author} {\bibfnamefont {B.~F.}\ \bibnamefont
  {Farrell}}\ and\ \bibinfo {author} {\bibfnamefont {P.~J.}\ \bibnamefont
  {Ioannou}},\ }\bibfield  {title} {\bibinfo {title} {A stochastic structural
  stability theory model of the drift wave-zonal flow system},\ }\href
  {https://doi.org/10.1063/1.3258666} {\bibfield  {journal} {\bibinfo
  {journal} {Phys. Plasmas}\ }\textbf {\bibinfo {volume} {16}},\ \bibinfo
  {pages} {112903} (\bibinfo {year} {2009}{\natexlab{b}})}\BibitemShut
  {NoStop}%
\bibitem [{\citenamefont {Farrell}\ and\ \citenamefont
  {Ioannou}(2017)}]{Farrell-Ioannou-2017-Saturn}%
  \BibitemOpen
  \bibfield  {author} {\bibinfo {author} {\bibfnamefont {B.~F.}\ \bibnamefont
  {Farrell}}\ and\ \bibinfo {author} {\bibfnamefont {P.~J.}\ \bibnamefont
  {Ioannou}},\ }\bibfield  {title} {\bibinfo {title} {Statistical state
  dynamics based theory for the formation and equilibration of {Saturn's} north
  polar jet},\ }\href {https://doi.org/10.1103/PhysRevFluids.2.073801}
  {\bibfield  {journal} {\bibinfo  {journal} {Phys. Rev. Fluids}\ }\textbf
  {\bibinfo {volume} {2}},\ \bibinfo {pages} {073801} (\bibinfo {year}
  {2017})}\BibitemShut {NoStop}%
\bibitem [{\citenamefont {Parker}\ and\ \citenamefont
  {Krommes}(2013)}]{Parker-Krommes-2013}%
  \BibitemOpen
  \bibfield  {author} {\bibinfo {author} {\bibfnamefont {J.~B.}\ \bibnamefont
  {Parker}}\ and\ \bibinfo {author} {\bibfnamefont {J.~A.}\ \bibnamefont
  {Krommes}},\ }\bibfield  {title} {\bibinfo {title} {Zonal flow as pattern
  formation},\ }\href {https://doi.org/10.1063/1.4828717} {\bibfield  {journal}
  {\bibinfo  {journal} {Phys. Plasmas}\ }\textbf {\bibinfo {volume} {20}},\
  \bibinfo {pages} {100703} (\bibinfo {year} {2013})}\BibitemShut {NoStop}%
\bibitem [{\citenamefont {Parker}\ and\ \citenamefont
  {Krommes}(2014)}]{Parker-Krommes-2014-generation}%
  \BibitemOpen
  \bibfield  {author} {\bibinfo {author} {\bibfnamefont {J.~B.}\ \bibnamefont
  {Parker}}\ and\ \bibinfo {author} {\bibfnamefont {J.~A.}\ \bibnamefont
  {Krommes}},\ }\bibfield  {title} {\bibinfo {title} {Generation of zonal flows
  through symmetry breaking of statistical homogeneity},\ }\href
  {https://doi.org/10.1088/1367-2630/16/3/035006} {\bibfield  {journal}
  {\bibinfo  {journal} {New J. Phys.}\ }\textbf {\bibinfo {volume} {16}},\
  \bibinfo {pages} {035006} (\bibinfo {year} {2014})}\BibitemShut {NoStop}%
\bibitem [{\citenamefont {Bakas}\ and\ \citenamefont
  {Ioannou}(2011)}]{Bakas-Ioannou-2011}%
  \BibitemOpen
  \bibfield  {author} {\bibinfo {author} {\bibfnamefont {N.~A.}\ \bibnamefont
  {Bakas}}\ and\ \bibinfo {author} {\bibfnamefont {P.~J.}\ \bibnamefont
  {Ioannou}},\ }\bibfield  {title} {\bibinfo {title} {Structural stability
  theory of two-dimensional fluid flow under stochastic forcing},\ }\href
  {https://doi.org/10.1017/jfm.2011.228} {\bibfield  {journal} {\bibinfo
  {journal} {J. Fluid Mech.}\ }\textbf {\bibinfo {volume} {682}},\ \bibinfo
  {pages} {332} (\bibinfo {year} {2011})}\BibitemShut {NoStop}%
\bibitem [{\citenamefont {Bakas}\ and\ \citenamefont
  {Ioannou}(2014)}]{Bakas-Ioannou-2014-jfm}%
  \BibitemOpen
  \bibfield  {author} {\bibinfo {author} {\bibfnamefont {N.~A.}\ \bibnamefont
  {Bakas}}\ and\ \bibinfo {author} {\bibfnamefont {P.~J.}\ \bibnamefont
  {Ioannou}},\ }\bibfield  {title} {\bibinfo {title} {A theory for the
  emergence of coherent structures in beta-plane turbulence},\ }\href
  {https://doi.org/10.1017/jfm.2013.663} {\bibfield  {journal} {\bibinfo
  {journal} {J. Fluid Mech.}\ }\textbf {\bibinfo {volume} {740}},\ \bibinfo
  {pages} {312} (\bibinfo {year} {2014})}\BibitemShut {NoStop}%
\bibitem [{\citenamefont {Constantinou}\ \emph {et~al.}(2016)\citenamefont
  {Constantinou}, \citenamefont {Farrell},\ and\ \citenamefont
  {Ioannou}}]{Constantinou-etal-2016}%
  \BibitemOpen
  \bibfield  {author} {\bibinfo {author} {\bibfnamefont {N.~C.}\ \bibnamefont
  {Constantinou}}, \bibinfo {author} {\bibfnamefont {B.~F.}\ \bibnamefont
  {Farrell}},\ and\ \bibinfo {author} {\bibfnamefont {P.~J.}\ \bibnamefont
  {Ioannou}},\ }\bibfield  {title} {\bibinfo {title} {Statistical state
  dynamics of jet--wave coexistence in barotropic beta-plane turbulence},\
  }\href {https://doi.org/10.1175/JAS-D-15-0288.1} {\bibfield  {journal}
  {\bibinfo  {journal} {J. Atmos. Sci.}\ }\textbf {\bibinfo {volume} {73}},\
  \bibinfo {pages} {2229} (\bibinfo {year} {2016})}\BibitemShut {NoStop}%
\bibitem [{Met()}]{Methods-Supplement}%
  \BibitemOpen
  \href@noop {} {}\bibinfo {note} {See Supplemental Material at [URL] for the
  numerical methods and the integration parameters, which includes
  Ref.~\cite{Kim-etal-1987}.}\BibitemShut {Stop}%
\bibitem [{\citenamefont {Hamilton}\ \emph {et~al.}(1995)\citenamefont
  {Hamilton}, \citenamefont {Kim},\ and\ \citenamefont
  {Waleffe}}]{Hamilton-etal-1995}%
  \BibitemOpen
  \bibfield  {author} {\bibinfo {author} {\bibfnamefont {J.~M.}\ \bibnamefont
  {Hamilton}}, \bibinfo {author} {\bibfnamefont {J.}~\bibnamefont {Kim}},\ and\
  \bibinfo {author} {\bibfnamefont {F.}~\bibnamefont {Waleffe}},\ }\bibfield
  {title} {\bibinfo {title} {Regeneration mechanisms of near-wall turbulence
  structures},\ }\href {https://doi.org/10.1017/S0022112095000978} {\bibfield
  {journal} {\bibinfo  {journal} {J. Fluid Mech.}\ }\textbf {\bibinfo {volume}
  {287}},\ \bibinfo {pages} {317} (\bibinfo {year} {1995})}\BibitemShut
  {NoStop}%
\bibitem [{\citenamefont {Farrell}\ \emph {et~al.}(2017)\citenamefont
  {Farrell}, \citenamefont {Ioannou},\ and\ \citenamefont
  {Nikolaidis}}]{Farrell-Ioannou-2017-bifur}%
  \BibitemOpen
  \bibfield  {author} {\bibinfo {author} {\bibfnamefont {B.~F.}\ \bibnamefont
  {Farrell}}, \bibinfo {author} {\bibfnamefont {P.~J.}\ \bibnamefont
  {Ioannou}},\ and\ \bibinfo {author} {\bibfnamefont {M.-A.}\ \bibnamefont
  {Nikolaidis}},\ }\bibfield  {title} {\bibinfo {title} {Instability of the
  roll--streak structure induced by background turbulence in pre-transitional
  {Couette} flow},\ }\href {https://doi.org/10.1103/PhysRevFluids.2.034607}
  {\bibfield  {journal} {\bibinfo  {journal} {Phys. Rev. Fluids}\ }\textbf
  {\bibinfo {volume} {2}},\ \bibinfo {pages} {034607} (\bibinfo {year}
  {2017})}\BibitemShut {NoStop}%
\bibitem [{\citenamefont {Nikolaidis}(2024)}]{Nikolaidis-2024}%
  \BibitemOpen
  \bibfield  {author} {\bibinfo {author} {\bibfnamefont {M.-A.}\ \bibnamefont
  {Nikolaidis}},\ }\href {https://arXiv.org/abs/2306.12739} {\bibinfo {title}
  {A vectorized {Navier-Stokes} ensemble direct numerical simulation code for
  plane parallel flows}} (\bibinfo {year} {2024}),\ \Eprint
  {https://arxiv.org/abs/2306.12739} {arXiv:2306.12739} \BibitemShut {NoStop}%
\bibitem [{Com()}]{Comparison-Supplement}%
  \BibitemOpen
  \href@noop {} {}\bibinfo {note} {See Supplemental Material at [URL] for the
  comparison between the S3T stable equilibrium and the corresponding DNS
  turbulent state, which includes
  Refs.~\cite{Bretheim-etal-2015,Farrell-etal-2016-VLSM,Bretheim-etal-2018}.}\BibitemShut
  {Stop}%
\bibitem [{\citenamefont {Malkus}(1956)}]{Malkus-1956}%
  \BibitemOpen
  \bibfield  {author} {\bibinfo {author} {\bibfnamefont {W.~V.~R.}\
  \bibnamefont {Malkus}},\ }\bibfield  {title} {\bibinfo {title} {Outline of a
  theory of turbulent shear flow},\ }\href
  {https://doi.org/10.1017/S0022112056000342} {\bibfield  {journal} {\bibinfo
  {journal} {J. Fluid Mech.}\ }\textbf {\bibinfo {volume} {1}},\ \bibinfo
  {pages} {521} (\bibinfo {year} {1956})}\BibitemShut {NoStop}%
\bibitem [{\citenamefont {Reynolds}\ and\ \citenamefont
  {Tiederman}(1967)}]{Reynolds-Tiederman-1967}%
  \BibitemOpen
  \bibfield  {author} {\bibinfo {author} {\bibfnamefont {W.~C.}\ \bibnamefont
  {Reynolds}}\ and\ \bibinfo {author} {\bibfnamefont {W.~G.}\ \bibnamefont
  {Tiederman}},\ }\bibfield  {title} {\bibinfo {title} {Stability of turbulent
  channel flow, with application to {Malkus's} theory},\ }\href
  {https://doi.org/10.1017/S0022112067000308} {\bibfield  {journal} {\bibinfo
  {journal} {J. Fluid Mech.}\ }\textbf {\bibinfo {volume} {27}},\ \bibinfo
  {pages} {253} (\bibinfo {year} {1967})}\BibitemShut {NoStop}%
\bibitem [{\citenamefont {Townsend}(1961)}]{Townsend-1961}%
  \BibitemOpen
  \bibfield  {author} {\bibinfo {author} {\bibfnamefont {A.~A.}\ \bibnamefont
  {Townsend}},\ }\bibfield  {title} {\bibinfo {title} {Equilibrium layers and
  wall turbulence},\ }\href {https://doi.org/10.1017/S0022112061000883}
  {\bibfield  {journal} {\bibinfo  {journal} {Journal of Fluid Mechanics}\
  }\textbf {\bibinfo {volume} {11}},\ \bibinfo {pages} {97–120} (\bibinfo
  {year} {1961})}\BibitemShut {NoStop}%
\bibitem [{\citenamefont {Nagata}(1990)}]{Nagata-1990}%
  \BibitemOpen
  \bibfield  {author} {\bibinfo {author} {\bibfnamefont {M.}~\bibnamefont
  {Nagata}},\ }\bibfield  {title} {\bibinfo {title} {Three-dimensional
  finite-amplitude solutions in plane {Couette} flow: bifurcation from
  infinity},\ }\href {https://doi.org/10.1017/S0022112090000829} {\bibfield
  {journal} {\bibinfo  {journal} {J. Fluid Mech.}\ }\textbf {\bibinfo {volume}
  {217}},\ \bibinfo {pages} {519} (\bibinfo {year} {1990})}\BibitemShut
  {NoStop}%
\bibitem [{\citenamefont {Waleffe}(1998)}]{Waleffe-1998}%
  \BibitemOpen
  \bibfield  {author} {\bibinfo {author} {\bibfnamefont {F.}~\bibnamefont
  {Waleffe}},\ }\bibfield  {title} {\bibinfo {title} {Three-dimensional
  coherent states in plane shear flows},\ }\href
  {https://doi.org/10.1103/PhysRevLett.81.4140} {\bibfield  {journal} {\bibinfo
   {journal} {Phys. Rev. Lett.}\ }\textbf {\bibinfo {volume} {81}},\ \bibinfo
  {pages} {4140} (\bibinfo {year} {1998})}\BibitemShut {NoStop}%
\bibitem [{\citenamefont {Waleffe}(2001)}]{Waleffe-2001}%
  \BibitemOpen
  \bibfield  {author} {\bibinfo {author} {\bibfnamefont {F.}~\bibnamefont
  {Waleffe}},\ }\bibfield  {title} {\bibinfo {title} {Exact coherent structures
  in channel flow},\ }\href {https://doi.org/10.1017/S0022112001004189}
  {\bibfield  {journal} {\bibinfo  {journal} {J. Fluid Mech.}\ }\textbf
  {\bibinfo {volume} {435}},\ \bibinfo {pages} {93} (\bibinfo {year}
  {2001})}\BibitemShut {NoStop}%
\bibitem [{\citenamefont {Waleffe}(2003)}]{Waleffe-2003}%
  \BibitemOpen
  \bibfield  {author} {\bibinfo {author} {\bibfnamefont {F.}~\bibnamefont
  {Waleffe}},\ }\bibfield  {title} {\bibinfo {title} {Homotopy of exact
  coherent structures in plane shear flows},\ }\href
  {https://doi.org/10.1063/1.1566753} {\bibfield  {journal} {\bibinfo
  {journal} {Phys. Fluids}\ }\textbf {\bibinfo {volume} {15}},\ \bibinfo
  {pages} {1517} (\bibinfo {year} {2003})}\BibitemShut {NoStop}%
\bibitem [{\citenamefont {Wang}\ \emph {et~al.}(2007)\citenamefont {Wang},
  \citenamefont {Gibson},\ and\ \citenamefont {Waleffe}}]{Wang-etal-2007}%
  \BibitemOpen
  \bibfield  {author} {\bibinfo {author} {\bibfnamefont {J.}~\bibnamefont
  {Wang}}, \bibinfo {author} {\bibfnamefont {J.}~\bibnamefont {Gibson}},\ and\
  \bibinfo {author} {\bibfnamefont {F.}~\bibnamefont {Waleffe}},\ }\bibfield
  {title} {\bibinfo {title} {Lower branch coherent states in shear flows:
  {Transition} and control},\ }\href
  {https://doi.org/10.1103/PhysRevLett.98.204501} {\bibfield  {journal}
  {\bibinfo  {journal} {Phys. Rev. Lett.}\ }\textbf {\bibinfo {volume} {98}},\
  \bibinfo {pages} {204501} (\bibinfo {year} {2007})}\BibitemShut {NoStop}%
\bibitem [{\citenamefont {Viswanath}(2007)}]{Viswanath-2007}%
  \BibitemOpen
  \bibfield  {author} {\bibinfo {author} {\bibfnamefont {D.}~\bibnamefont
  {Viswanath}},\ }\bibfield  {title} {\bibinfo {title} {Recurrent motions
  within plane {Couette} turbulence},\ }\href
  {https://doi.org/10.1017/S0022112007005459} {\bibfield  {journal} {\bibinfo
  {journal} {Journal of Fluid Mechanics}\ }\textbf {\bibinfo {volume} {580}},\
  \bibinfo {pages} {339–358} (\bibinfo {year} {2007})}\BibitemShut {NoStop}%
\bibitem [{\citenamefont {Gibson}\ \emph {et~al.}(2009)\citenamefont {Gibson},
  \citenamefont {Halcrow},\ and\ \citenamefont
  {Cvitanovi\'c}}]{Gibson-etal-2009}%
  \BibitemOpen
  \bibfield  {author} {\bibinfo {author} {\bibfnamefont {J.~F.}\ \bibnamefont
  {Gibson}}, \bibinfo {author} {\bibfnamefont {J.}~\bibnamefont {Halcrow}},\
  and\ \bibinfo {author} {\bibfnamefont {P.}~\bibnamefont {Cvitanovi\'c}},\
  }\bibfield  {title} {\bibinfo {title} {Equilibrium and travelling-wave
  solutions of plane {C}ouette flow},\ }\href@noop {} {\bibfield  {journal}
  {\bibinfo  {journal} {J. Fluid Mech.}\ }\textbf {\bibinfo {volume} {638}},\
  \bibinfo {pages} {243} (\bibinfo {year} {2009})}\BibitemShut {NoStop}%
\bibitem [{\citenamefont {Hall}\ and\ \citenamefont
  {Sherwin}(2010)}]{Hall-Sherwin-2010}%
  \BibitemOpen
  \bibfield  {author} {\bibinfo {author} {\bibfnamefont {P.}~\bibnamefont
  {Hall}}\ and\ \bibinfo {author} {\bibfnamefont {S.}~\bibnamefont {Sherwin}},\
  }\bibfield  {title} {\bibinfo {title} {Streamwise vortices in shear flows:
  harbingers of transition and the skeleton of coherent structures},\ }\href
  {https://doi.org/10.1017/S0022112010002892} {\bibfield  {journal} {\bibinfo
  {journal} {Journal of Fluid Mechanics}\ }\textbf {\bibinfo {volume} {661}},\
  \bibinfo {pages} {178–205} (\bibinfo {year} {2010})}\BibitemShut {NoStop}%
\bibitem [{\citenamefont {Blackburn}\ \emph {et~al.}(2013)\citenamefont
  {Blackburn}, \citenamefont {Hall},\ and\ \citenamefont
  {Sherwin}}]{Blackburn-Hall-2013}%
  \BibitemOpen
  \bibfield  {author} {\bibinfo {author} {\bibfnamefont {H.~M.}\ \bibnamefont
  {Blackburn}}, \bibinfo {author} {\bibfnamefont {P.}~\bibnamefont {Hall}},\
  and\ \bibinfo {author} {\bibfnamefont {S.~J.}\ \bibnamefont {Sherwin}},\
  }\bibfield  {title} {\bibinfo {title} {Lower branch equilibria in {Couette}
  flow: the emergence of canonical states for arbitrary shear flows},\ }\href
  {https://doi.org/10.1017/jfm.2013.254} {\bibfield  {journal} {\bibinfo
  {journal} {Journal of Fluid Mechanics}\ }\textbf {\bibinfo {volume} {726}},\
  \bibinfo {pages} {R2} (\bibinfo {year} {2013})}\BibitemShut {NoStop}%
\bibitem [{\citenamefont {Deguchi}\ \emph {et~al.}(2013)\citenamefont
  {Deguchi}, \citenamefont {Hall},\ and\ \citenamefont
  {Walton}}]{Deguchi-etal-2013}%
  \BibitemOpen
  \bibfield  {author} {\bibinfo {author} {\bibfnamefont {K.}~\bibnamefont
  {Deguchi}}, \bibinfo {author} {\bibfnamefont {P.}~\bibnamefont {Hall}},\ and\
  \bibinfo {author} {\bibfnamefont {A.}~\bibnamefont {Walton}},\ }\bibfield
  {title} {\bibinfo {title} {The emergence of localized vortex–wave
  interaction states in plane {Couette} flow},\ }\href
  {https://doi.org/10.1017/jfm.2013.27} {\bibfield  {journal} {\bibinfo
  {journal} {Journal of Fluid Mechanics}\ }\textbf {\bibinfo {volume} {721}},\
  \bibinfo {pages} {58–85} (\bibinfo {year} {2013})}\BibitemShut {NoStop}%
\bibitem [{\citenamefont {Deguchi}\ and\ \citenamefont
  {Hall}(2014)}]{Deguchi-Hall-2014}%
  \BibitemOpen
  \bibfield  {author} {\bibinfo {author} {\bibfnamefont {K.}~\bibnamefont
  {Deguchi}}\ and\ \bibinfo {author} {\bibfnamefont {P.}~\bibnamefont {Hall}},\
  }\bibfield  {title} {\bibinfo {title} {The high-{R}eynolds-number asymptotic
  development of nonlinear equilibrium states in plane {Couette} flow},\ }\href
  {https://doi.org/10.1017/jfm.2014.234} {\bibfield  {journal} {\bibinfo
  {journal} {Journal of Fluid Mechanics}\ }\textbf {\bibinfo {volume} {750}},\
  \bibinfo {pages} {99–112} (\bibinfo {year} {2014})}\BibitemShut {NoStop}%
\bibitem [{Sca()}]{Scaling-Supplement}%
  \BibitemOpen
  \href@noop {} {}\bibinfo {note} {See Supplemental Material at [URL] for the
  scaling argument demonstrating that roll advection dominates viscous
  diffusion in determining the structure of the fluctuation eigenmodes, which
  includes Refs.~\cite{Drazin-Reid-81,Schmid-Henningson-2001}.}\BibitemShut
  {Stop}%
\bibitem [{\citenamefont {Jiménez}\ \emph {et~al.}(2005)\citenamefont
  {Jiménez}, \citenamefont {Kawahara}, \citenamefont {Simens}, \citenamefont
  {Nagata},\ and\ \citenamefont {Shiba}}]{Jimenez-Kawahara-2005}%
  \BibitemOpen
  \bibfield  {author} {\bibinfo {author} {\bibfnamefont {J.}~\bibnamefont
  {Jiménez}}, \bibinfo {author} {\bibfnamefont {G.}~\bibnamefont {Kawahara}},
  \bibinfo {author} {\bibfnamefont {M.~P.}\ \bibnamefont {Simens}}, \bibinfo
  {author} {\bibfnamefont {M.}~\bibnamefont {Nagata}},\ and\ \bibinfo {author}
  {\bibfnamefont {M.}~\bibnamefont {Shiba}},\ }\bibfield  {title} {\bibinfo
  {title} {Characterization of near-wall turbulence in terms of equilibrium and
  “bursting” solutions},\ }\href {https://doi.org/10.1063/1.1825451}
  {\bibfield  {journal} {\bibinfo  {journal} {Physics of Fluids}\ }\textbf
  {\bibinfo {volume} {17}},\ \bibinfo {pages} {015105} (\bibinfo {year}
  {2005})}\BibitemShut {NoStop}%
\bibitem [{\citenamefont {Kawahara}(2009)}]{Kawahara-2009}%
  \BibitemOpen
  \bibfield  {author} {\bibinfo {author} {\bibfnamefont {G.}~\bibnamefont
  {Kawahara}},\ }\bibfield  {title} {\bibinfo {title} {Theoretical
  interpretation of coherent structures in near-wall turbulence},\ }\href@noop
  {} {\bibfield  {journal} {\bibinfo  {journal} {Fluid Dyn. Res.}\ }\textbf
  {\bibinfo {volume} {41}},\ \bibinfo {pages} {064001} (\bibinfo {year}
  {2009})}\BibitemShut {NoStop}%
\bibitem [{\citenamefont {Kawahara}\ \emph {et~al.}(2012)\citenamefont
  {Kawahara}, \citenamefont {Uhlmann},\ and\ \citenamefont
  {Van~Veen}}]{Kawahara-etal-2012}%
  \BibitemOpen
  \bibfield  {author} {\bibinfo {author} {\bibfnamefont {G.}~\bibnamefont
  {Kawahara}}, \bibinfo {author} {\bibfnamefont {M.}~\bibnamefont {Uhlmann}},\
  and\ \bibinfo {author} {\bibfnamefont {L.}~\bibnamefont {Van~Veen}},\
  }\bibfield  {title} {\bibinfo {title} {The significance of simple invariant
  solutions in turbulent flows},\ }\href@noop {} {\bibfield  {journal}
  {\bibinfo  {journal} {{Annu. Rev. Fluid Mech.}}\ }\textbf {\bibinfo {volume}
  {44}},\ \bibinfo {pages} {203} (\bibinfo {year} {2012})}\BibitemShut
  {NoStop}%
\bibitem [{\citenamefont {Farrell}\ and\ \citenamefont
  {Ioannou}(2026)}]{Farrell-Ioannou-2026}%
  \BibitemOpen
  \bibfield  {author} {\bibinfo {author} {\bibfnamefont {B.~F.}\ \bibnamefont
  {Farrell}}\ and\ \bibinfo {author} {\bibfnamefont {P.~J.}\ \bibnamefont
  {Ioannou}},\ }\bibfield  {title} {\bibinfo {title} {Statistical state
  dynamics modes and equilibria underlie the structure and mechanism of wide
  channel {Couette} turbulence},\ }\href
  {https://doi.org/10.1088/1742-6596/3230/1/012020} {\bibfield  {journal}
  {\bibinfo  {journal} {J. Phys. Conf. Ser.}\ }\textbf {\bibinfo {volume}
  {3230}},\ \bibinfo {pages} {012020} (\bibinfo {year} {2026})}\BibitemShut
  {NoStop}%
\bibitem [{\citenamefont {Kim}\ \emph {et~al.}(1987)\citenamefont {Kim},
  \citenamefont {Moin},\ and\ \citenamefont {Moser}}]{Kim-etal-1987}%
  \BibitemOpen
  \bibfield  {author} {\bibinfo {author} {\bibfnamefont {J.}~\bibnamefont
  {Kim}}, \bibinfo {author} {\bibfnamefont {P.}~\bibnamefont {Moin}},\ and\
  \bibinfo {author} {\bibfnamefont {R.}~\bibnamefont {Moser}},\ }\bibfield
  {title} {\bibinfo {title} {Turbulence statistics in fully developed channel
  flow at low {Reynolds} number},\ }\href
  {https://doi.org/10.1017/S0022112087000892} {\bibfield  {journal} {\bibinfo
  {journal} {J. Fluid Mech.}\ }\textbf {\bibinfo {volume} {177}},\ \bibinfo
  {pages} {133} (\bibinfo {year} {1987})}\BibitemShut {NoStop}%
\bibitem [{\citenamefont {Bretheim}\ \emph {et~al.}(2015)\citenamefont
  {Bretheim}, \citenamefont {Meneveau},\ and\ \citenamefont
  {Gayme}}]{Bretheim-etal-2015}%
  \BibitemOpen
  \bibfield  {author} {\bibinfo {author} {\bibfnamefont {J.~U.}\ \bibnamefont
  {Bretheim}}, \bibinfo {author} {\bibfnamefont {C.}~\bibnamefont {Meneveau}},\
  and\ \bibinfo {author} {\bibfnamefont {D.~F.}\ \bibnamefont {Gayme}},\
  }\bibfield  {title} {\bibinfo {title} {Standard logarithmic mean velocity
  distribution in a band-limited restricted nonlinear model of turbulent flow
  in a half-channel},\ }\href {https://doi.org/10.1063/1.4906987} {\bibfield
  {journal} {\bibinfo  {journal} {Phys. Fluids}\ }\textbf {\bibinfo {volume}
  {27}},\ \bibinfo {pages} {011702} (\bibinfo {year} {2015})}\BibitemShut
  {NoStop}%
\bibitem [{\citenamefont {Farrell}\ \emph {et~al.}(2016)\citenamefont
  {Farrell}, \citenamefont {Ioannou}, \citenamefont {Jim{\'e}nez},
  \citenamefont {Constantinou}, \citenamefont {Lozano-Dur{\'a}n},\ and\
  \citenamefont {Nikolaidis}}]{Farrell-etal-2016-VLSM}%
  \BibitemOpen
  \bibfield  {author} {\bibinfo {author} {\bibfnamefont {B.~F.}\ \bibnamefont
  {Farrell}}, \bibinfo {author} {\bibfnamefont {P.~J.}\ \bibnamefont
  {Ioannou}}, \bibinfo {author} {\bibfnamefont {J.}~\bibnamefont
  {Jim{\'e}nez}}, \bibinfo {author} {\bibfnamefont {N.~C.}\ \bibnamefont
  {Constantinou}}, \bibinfo {author} {\bibfnamefont {A.}~\bibnamefont
  {Lozano-Dur{\'a}n}},\ and\ \bibinfo {author} {\bibfnamefont {M.-A.}\
  \bibnamefont {Nikolaidis}},\ }\bibfield  {title} {\bibinfo {title} {A
  statistical state dynamics-based study of the structure and mechanism of
  large-scale motions in plane {Poiseuille} flow},\ }\href
  {https://doi.org/10.1017/jfm.2016.661} {\bibfield  {journal} {\bibinfo
  {journal} {J. Fluid Mech.}\ }\textbf {\bibinfo {volume} {809}},\ \bibinfo
  {pages} {290} (\bibinfo {year} {2016})}\BibitemShut {NoStop}%
\bibitem [{\citenamefont {Bretheim}\ \emph {et~al.}(2018)\citenamefont
  {Bretheim}, \citenamefont {Meneveau},\ and\ \citenamefont
  {Gayme}}]{Bretheim-etal-2018}%
  \BibitemOpen
  \bibfield  {author} {\bibinfo {author} {\bibfnamefont {J.~U.}\ \bibnamefont
  {Bretheim}}, \bibinfo {author} {\bibfnamefont {C.}~\bibnamefont {Meneveau}},\
  and\ \bibinfo {author} {\bibfnamefont {D.~F.}\ \bibnamefont {Gayme}},\
  }\bibfield  {title} {\bibinfo {title} {A restricted nonlinear large eddy
  simulation model for high {Reynolds} number flows},\ }\href
  {https://doi.org/10.1080/14685248.2017.1403031} {\bibfield  {journal}
  {\bibinfo  {journal} {Journal of {Turbulence}}\ }\textbf {\bibinfo {volume}
  {19}},\ \bibinfo {pages} {141} (\bibinfo {year} {2018})}\BibitemShut
  {NoStop}%
\bibitem [{\citenamefont {Drazin}\ and\ \citenamefont
  {Reid}(1981)}]{Drazin-Reid-81}%
  \BibitemOpen
  \bibfield  {author} {\bibinfo {author} {\bibfnamefont {P.~G.}\ \bibnamefont
  {Drazin}}\ and\ \bibinfo {author} {\bibfnamefont {W.~H.}\ \bibnamefont
  {Reid}},\ }\href@noop {} {\emph {\bibinfo {title} {Hydrodynamic Stability}}}\
  (\bibinfo  {publisher} {Cambridge University Press, Cambridge},\ \bibinfo
  {year} {1981})\BibitemShut {NoStop}%
\bibitem [{\citenamefont {Schmid}\ and\ \citenamefont
  {Henningson}(2001)}]{Schmid-Henningson-2001}%
  \BibitemOpen
  \bibfield  {author} {\bibinfo {author} {\bibfnamefont {P.~J.}\ \bibnamefont
  {Schmid}}\ and\ \bibinfo {author} {\bibfnamefont {D.~S.}\ \bibnamefont
  {Henningson}},\ }\href@noop {} {\emph {\bibinfo {title} {Stability and
  Transition in Shear Flows}}}\ (\bibinfo  {publisher} {Springer, New York},\
  \bibinfo {year} {2001})\BibitemShut {NoStop}%
\end{thebibliography}%

\end{document}